# How does Object-Oriented Code Refactoring Influence Software Quality? Research Landscape and Challenges

**Satnam Kaur** [*1], **Paramvir Singh**[2]

[1]Department of Computer Science and Engineering

Dr B R Ambedkar National Institute of Technology, Jalandhar 144011, Punjab, India

[2] School of Computer Science

University of Auckland, Auckland 1142, New Zealand

## ABSTRACT

**Context:** Software refactoring aims to improve software quality and developer productivity. Numerous empirical studies investigating the impact of refactoring activities on software quality have been conducted over the last two decades.

**Objective:** This study aims to perform a comprehensive systematic mapping study of existing empirical studies on evaluation of the effect of object-oriented code refactoring activities on software quality attributes.

**Method:** We followed a multi-stage scrutinizing process to select 142 primary studies published till December 2017. The selected primary studies were further classified based on several aspects to answer the research questions defined for this work. In addition, we applied vote-counting approach to combine the empirical results and their analysis reported in primary studies.

**Results:** The findings indicate that studies conducted in academic settings found more positive impact of refactoring on software quality than studies performed in industries. In general, refactoring activities caused all quality attributes to improve or degrade except for cohesion, complexity, inheritance, fault-proneness and power consumption attributes. Furthermore, individual refactoring activities have variable effects on most quality attributes explored in primary studies, indicating that refactoring does not always improve all quality attributes.

**Conclusions:** This study points out several open issues which require further investigation, e.g., lack of industrial validation, lesser coverage of refactoring activities, limited tool support, etc.

## 1. INTRODUCTION

Software maintenance is considered to be one of the costliest and effort-intensive software development activities [1-2]. One of the prominent factors sourcing high maintenance costs is the bad software design quality [3]. During continuous maintenance of the software, various code smells are unintentionally introduced by the developers [4-5]. These code smells may degrade the quality of the software gradually over time [6-7]. Basically, code smell is a symptom in the source code that indicates a deeper problem which can be removed by the right choice of *refactoring activities* [8].

The term 'refactoring' was coined by Opdyke in 1992 [9] and defines a disciplined process of restructuring the source code in such a way that it improves the internal structure of the software while preserving its external behaviour [8]. Here behaviour preservation means that software must produce the same output before and after the application of a specific refactoring activity. Hence, refactoring intends to provide a solution for improving the quality of the software, which in turn reduces the maintenance cost. In object oriented software engineering community, it is widely believed that refactoring results in a better quality software through the eradication of bad smells, and also improves developer's productivity [8, 10].

However, there are some ambiguities regarding this notion too, with a number of studies claiming that refactoring degrades the quality of software [11-15]. In order to extract a clearer picture, we conducted a systematic mapping study that analyses and reports the empirical findings on the relationship between refactoring activities and quality of object-oriented software. This study, we believe, will help in reducing some of the ambiguities regarding the benefits and limitations of refactoring activities. In broader terms, the motivations behind our work can be stated as follows.

1. Various industry practitioners are oblivious to the impact of refactoring activities on software quality. Consequently, while performing any refactoring activity, they feel reluctant as they believe refactoring may either break the code or result in the wastage of resources [16-17]. Their concerns can be attributed to the scarcity of appropriate sources that discuss state-of-the-art of software refactoring process and its benefits in a lucid manner. Hence, this paper reports an in-depth systematic mapping study (SMS) to analyse, synthesize and present the empirical findings related to the effect of refactoring activities targeting code smells on software quality. Awareness about the effect of planned refactoring changes on software quality will boost the confidence of developers in exercising software refactoring as regular development practice.
2. Retrospectively, it is observed that so far there have been only four literature surveys covering the relationship between refactoring and software quality. Among them, three [18-20] partially and one comprehensively (Dallal and Abdin; published in 2017 covering the literature till 2015) [21] explored the impact of refactoring activities on object-oriented software quality. Nevertheless, the magnitude of research in this area has been proliferating continuously. One quick overview of the area reflected that many relevant studies were published in last two years (2016 and 2017) and hence were not included in Dallal and Abdin's SLR [21]. This prompted us to evaluate and interpret the ever-growing research in this field by following a rigorous systematic mapping process.

We initially identified a total of 13,283 potentially relevant articles, published in literature till December 2017, from six digital libraries and academic search engines. Later, we scrutinized these articles based on different perspectives, including title, abstract and full text, followed by manual search and reference snowballing. Furthermore, we also analysed the quality of these studies to select 142 Primary Studies (PSs). These selected PSs were further analysed in detail to make following major contributions:

1. We classify the selected PSs based on the number of features related to the impact of refactoring activities on software quality including research contribution, context, refactoring activities, focus, investigation approach, datasets, and software quality measures.
2. We discuss various refactoring activities, software quality measures, statistical techniques, quality attributes and datasets, employed by selected PSs.
3. We extract the reported software tools that predict or assess the impact of refactoring activities on software quality.
4. We report the current state-of-the-art of the existing empirical studies evaluating the effect of object-oriented code refactoring activities on software quality attributes.
5. Finally, we collate the dispersed and contradictory findings to derive a list of open issues and challenges needed to be addressed in future.

We believe that the outcomes of this SMS will help the software engineering researchers in identifying open areas where further research is required, as well as the practitioners who want to get an updated view of the current state of research in this field. This study will also assist the scientific community in comprehending the issues and challenges that must be addressed for determining the effectiveness of refactoring activities in developing high quality software product.

The remainder of this paper is structured as follows. Section 2 provides a brief primer on software refactoring followed by the discussion on most closely related SLRs in the domain of software refactoring. Section 3 elaborates the mapping study protocol followed in this work. Section 4 presents the overview of PSs along with answers to the research questions. The discussion on the research questions is provided in Section 5. Section 6 summarizes the potential threats to the validity of this work. Finally, Section 7 concludes this SMS and outlines the possible directions for future work.

## 2. BACKGROUND AND RELATED WORK

In this section, we describe the general refactoring process along with related literature reviews and surveys conducted in the area of software refactoring.

### 2.1. Refactoring process and stages

A typical refactoring process includes the following stages [18]:

a) Identify code locations where the software should be refactored.
b) Determine appropriate refactoring activities to be applied.
c) Guarantee that the applied refactoring activities preserve the software behaviour.
d) Apply the selected refactoring activities.
e) Assess the impact of refactoring activities on software quality.

f) Maintain the consistency between the refactored software and other software artifacts such as requirement specifications, design documents, test cases, documentation, etc.

The aforementioned refactoring stages are further divided into three phases, namely refactoring opportunity identification, refactoring application, and refactoring maintenance [22], as shown in Figure 1.

### 2.2. Related literature

The previous research related to software refactoring comprises a number of literature reviews and surveys. Table 1 presents a comparative summary of these secondary studies with our study highlighting their focus and limitations. The first extensive survey related to software refactoring was conducted by Mens and Tourwe [18] in 2004. It covered several aspects like refactoring process stages and their supporting techniques, software artifact types, refactoring tool support, and the effect of refactoring activities on software process. In contrast to this (non-systematic) survey, we conduct a SMS with a narrowed down scope of determining the impact of refactoring activities on software quality incorporating the ever-growing related research conducted between 2003 and 2017.

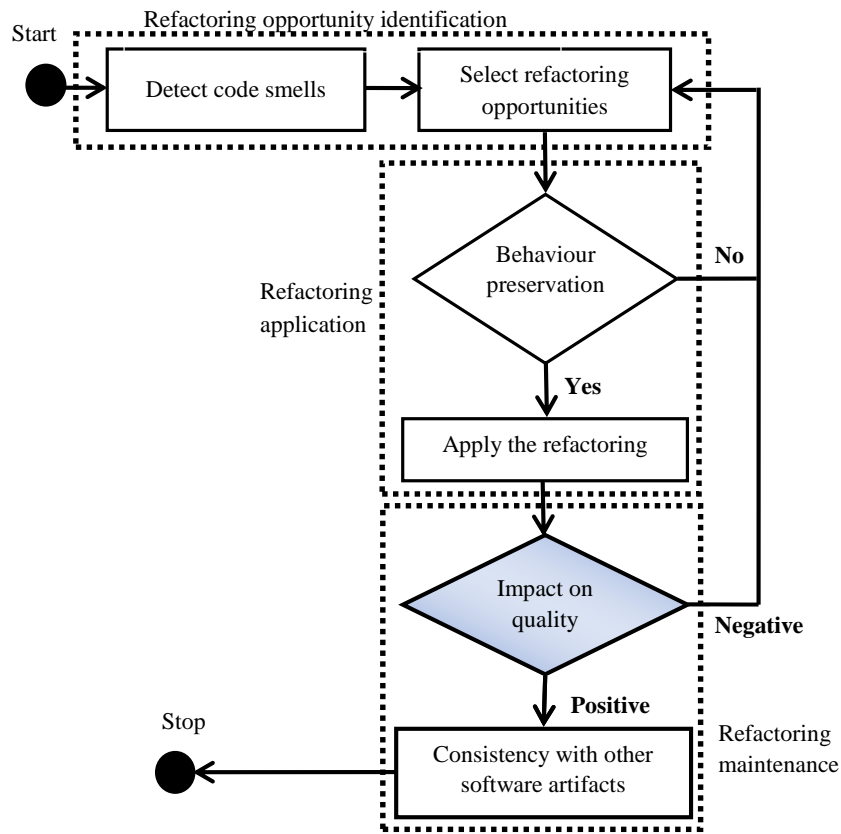

**Figure 1: Flow chart representation of refactoring steps and our study scope**

Wangberg [19] analysed 46 studies to describe state-of-the-art in bad smells and refactoring activities. He reviewed several aspects like refactoring opportunity identification, refactoring application, impact of code smells on software quality, effect of refactoring activities on code quality, and methods and tools used to detect and remove code smells. Wangberg's [19] SLR is wider in scope as he covered these aspects for non object-oriented code and design artifacts like Unified Modeling Language (UML) models as well. Also, they extracted literature published till 2009 only from three digital libraries, namely, ACM Digital Library, ISI Web of Knowledge and IEEE Xplore. We restricted the scope of our SMS to object-oriented code based refactoring activities only. Furthermore, we employed a more comprehensive set of digital libraries, along with a manual search, to cover all relevant literature published till December 2017.

Table 1: Comparison of previous refactoring surveys and reviews

| Study | Year | Study Type | Period | PSs | Search Type* M | Search Type* A | Electronic Libraries Selected | Focus | Limitations |
|---|---|---|---|---|---|---|---|---|---|
| Mens and Tourwe [18] | 2004 | Survey | till 2003 | - | ✓ | × | Not Applicable (*NA*) | – refactoring stages<br>– software artifact types<br>– refactoring tool support<br>– effect of refactoring on software process | – did not follow a formal review process.<br>– considered few studies related to refactoring impact on quality with no thorough discussion or comparison. |
| Wangberg [19] (Master thesis; may not be a peer-reviewed study) | 2010 | SLR | 2000-2009 | 46 | × | ✓ | ACM Digital Library, ISI Web of Knowledge, IEEE Xplore | – refactoring opportunity identification<br>– refactoring application<br>– refactoring impact on quality<br>– code smell effect on quality<br>– methods and tools to detect and remove code smells | – covered literature from only three digital libraries.<br>– considered only one PS that performed a small case study concerning the effect of refactoring on software quality. |
| Misbhauddin and Alshayeb [23] | 2013 | SLR | 2001-2012 | 94 | × | ✓ | ACM Digital Library, IEEE Xplore, Wiley Inter Science Journal Finder, Science Direct, ISI Web of Science, Compendex, Google Scholar, SpringerLink | – UML models<br>– formalisms used<br>– refactoring effect on model quality | – only focused on model quality. |
| Abebe and Yoo [20] | 2014 | SLR | 1999-2013 | 58 | × | ✓ | ACM Digital Library, IEEE Xplore, Scopus, Wiley Online Library, Science Direct, Web of Science, Compendex, CiteSeer, SpringerLink | – refactoring of code, design and test artifacts<br>– refactoring tools | – considered few studies related to refactoring impact on quality. |
| Dallal and Abdin [21] | 2017 | SLR | till 2015 | 76 | × | ✓ | ACM Digital Library, IEEE Xplore,, ISI Web of Science, Science Direct, Compendex and Inspec, SpringerLink, Scopus | – refactoring activities<br>– quality attributes and measures<br>– statistical approaches<br>– datasets used<br>– impact of code refactoring activities on software quality | – did not address the impact of refactoring activities on software process, performance and people quality attributes. |
| This SMS | NA | SMS | till 2017 | 142 | ✓ | ✓ | IEEE Xplore, Wiley Online Library, Science Direct, SpringerLink, ACM Digital Library, Scopus | – code smell and refactoring activities<br>– software quality attributes and measures<br>– statistical approaches<br>– datasets used<br>– refactoring impact assessing tools<br>– impact of code refactoring activities on software quality | – scope is limited to impact of code refactoring activities on software quality |

*M: Manual Search, A: Automatic Search*

Misbhauddin and Alshayeb [23] conducted an SLR covering 94 primary studies published in the area of UML model refactoring. They overviewed the UML models considered for refactoring, formalisms used, and the effect of refactoring activities on model quality. In contrast to their SLR, we focus on studies determining the impact of refactoring on code quality rather than model quality. Abebe and Yoo [20] conducted an SLR on 58 PSs focusing on trends, opportunities and challenges in the field of software refactoring in a broader scope. They covered the refactoring of design and testing artifacts apart from code artifacts. Their SLR had a wider scope and they examined only a few studies related to refactoring impact on quality.

Among other related reviews, Zhang et al. [24] studied 39 papers to analyse the most popular code smells, objectives of studies on code smells, methods followed to study code smells, and provided empirical evidence to support the notion that code smells have negative impact on source code of software systems. Al Dallal's [25] SLR examined 47 studies based on several criteria, including refactoring activities, refactoring opportunities identification techniques, approaches followed to empirically evaluate the identification techniques, and datasets used to evaluate the identification approaches in an object-oriented context. Rochimah et al. [26] performed a literature review of 20 studies to review the refactoring process at non-source code level. Rasool and Arshad [27] reported an SLR to categorize, compare and assess existing tools and techniques used to detect code smells. Mariani and Vergilio [28] conducted an SLR that investigated 71 PSs based on software artifact and their representations, refactoring activities, approaches exploited to preserve behaviour, refactoring application, metrics explored, consistency with other artifacts, refactoring sequencing, evaluation approaches, search based algorithms, datasets used and refactoring tools. Later in 2018, Mohan and Greer [29] performed almost a similar SLR of 55 PSs focusing on refactoring approaches, search based techniques, datasets used, metrics explored and refactoring tools. de Paulo Sobrinho et al. [30] conducted an SLR covering 351 primary studies to describe state-of-the-art in code smells between the years 1990 and 2017. They investigated several aspects including evolution of researchers' interest in code smells, different groups or people interested in code smells, diversity of different types of code smells, objectives, outcomes and material for experimental setup, and distribution of research among several venues. Their SLR had a wider scope and examined quite a limited set of studies related to refactoring impact on quality. As a result, they did not provide detailed descriptions of studies related to our phenomenon of interest.

On the final stages of conducting this SMS, we encountered a systematic review by Dallal and Abdin [21] that is by far closest to our work. They analysed 76 empirical studies investigating the effect of refactoring activities on software quality covering several aspects like software quality measures and attributes, refactoring activities, datasets, evaluation approaches and impact results. However, our work is different from the SLR reported by Dallal and Abdin [21] in following respects.

a) Among 142 identified PSs, we obtained 66 (46.5%) additional PSs along with the 76 PSs selected by Dallal and Abdin [21]. It may be attributed to following reasons.
   - We considered a generic search string (as suggested by Kitchenham and Charters [31]) and covered the literature till December 2017, whereas Dallal and Abdin [21] used a specific search string and covered the studies till December 2015. The research in the area of refactoring is steadily maturing, unleashing a substantial number of relevant additional PSs (29 PSs) conducted in the last two years.
   - The studies considered in Dallal and Abdin's SLR [21] focus on addressing the effect of refactoring activities on software product quality attributes only. However, the other significant quality attributes viz. process, people, and performance quality attributes were not taken into account for the same. In that respect, our SMS is more comprehensive as we include aforementioned types of quality attributes to have a better insight into impact of refactoring activities on software quality.
   - Finally, the empirical studies that reported a tool to assess or predict the effect of refactoring activities on software quality attributes are also included in this work, in contrast to Dallal and Abdin's SLR [21], where there was no specific attempt to include such studies.
b) We classified the selected PSs based on the applied research method and study settings, in contrast to Dallal and Abdin's SLR [21] where the authors did not categorize the PSs according to the aforementioned dimensions.
c) It is contemplated from Dallal and Abdin's SLR [21] that no clear distinctions were made among the findings of studies performed in industrial and academic settings. It tends to mask the research gap between academic and industrial settings as studies conducted in these environments may lead to different outcomes regarding the effect of refactoring on quality. To bridge this research gap, we provide lucid insights for identifying the differences between the findings generated from experimentation in academic and industrial settings, so that the outcomes of such works could converge to common and useful conclusions.

Based on the above mentioned major differentiating factors, we did not find it suitable to compare the findings of this work with that of Dallal and Abdin [21] in detail. However, for the readers' assistance, we still provide a summary of the differences between the findings of this work and aforementioned SLR [21] in Subsection 5.6.

## 3. REVIEW METHOD

An SLR aims at identifying, assessing and interpreting all the available research related to a particular phenomenon or area of interest, using a defined and defendable search strategy [31]. A SMS is a type of SLR that collects and classifies the existing research to provide a broader overview of a particular field [32]. This SMS targets to aggregate and analyse the empirical findings relevant to the impact of code refactoring activities on object-oriented software quality. Following the guidelines proposed by Kitchenham and Charters [31] along with Petersen et al. [32] and Biolchini et al. [33], we conducted this SMS in three main stages namely planning, conducting and reporting the findings. The planning phase comprises determining the need of SMS and defining a mapping study protocol. The rationale behind this SMS was presented in previous sections. The major steps of the mapping study protocol followed are depicted in Figure 2. The elaboration of these steps is presented in subsequent subsections.

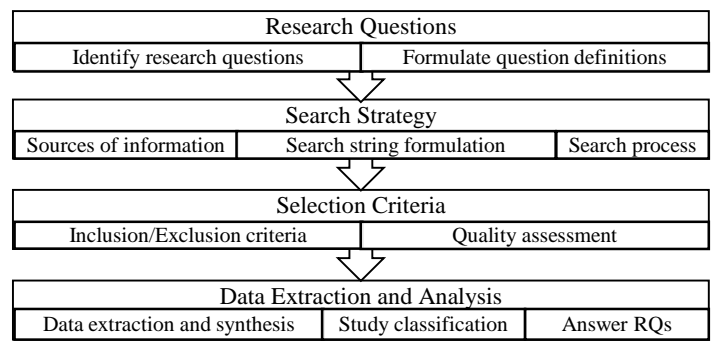

**Figure 2: Overview of mapping study protocol**

### 3.1 Research questions

Following the motivations drawn for this study, we defined following five research questions in order to capture various aspects of empirical studies on refactoring activities and software quality:

**RQ1.** Which refactoring activities and underlying code smells were investigated?

**RQ2.** Which software quality measures have been used to study the impact of refactoring activities on software quality?

**RQ3.** Which tools have been reported to predict or assess the impact of refactoring activities on software quality?

**RQ4.** What datasets were used to conduct the empirical studies investigating the impact of refactoring activities on software quality?

**RQ5.** What is the current state of knowledge about the impact of refactoring activities on software quality?

The mapping of the research questions to various steps involved in the process of assessing the effect of refactoring activities on software quality is shown in Figure 3. RQ1 focuses on identifying the most and least explored refactoring activities and underlying code smells among the selected PSs. This will help in determining the frequently handled code smells as well as the refactoring activities that are yet to be investigated in future research. RQ2 aims at providing a list of software quality measures in which the software researchers and developers working in the area of software refactoring, are profoundly interested.

Although existing research provided tools for the identification of refactoring opportunities or the automatic or semi-automatic applications of refactorings or even for their prioritization, most of these tools do not assess or predict the impact of refactoring activities on software quality. Therefore, without understanding the effect of a change caused by a particular refactoring in software quality, schedule-bound industry practitioners feel reluctant to adopt refactoring because of perceived cost, effort and time intensive outcome of refactoring [16, 34]. To address these concerns, RQ3 targets such tools that provide the prediction or assessment of the benefits of applying refactoring activities. A wide range of datasets have been used to empirically validate the impact of refactoring activities on software quality. RQ4 attempts to identify and classify the most explored datasets that have marked their presence in existing literature as well as opportunities for research targeting less explored programming languages or dataset types. RQ5 will assist in building an understanding of the reported benefits and drawbacks of applying software refactoring activities on desired software quality attributes. It targets to compile a body of knowledge that helps researchers and practitioners in being more informed about the impact of overall as well as individual refactoring activities on software quality. This will further help in finding specific challenges that need to be addressed in this area.

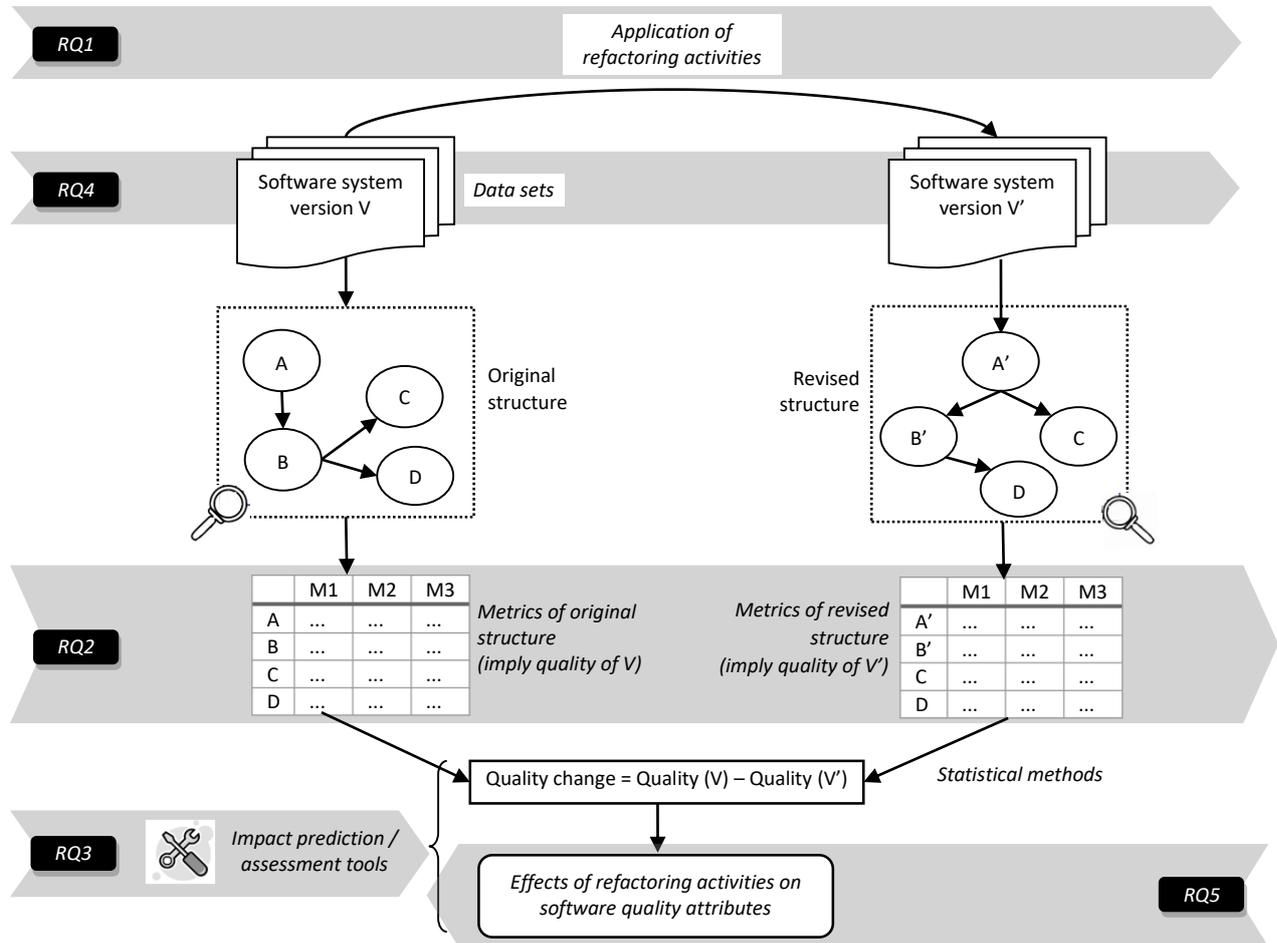

**Figure 3: Mapping of research questions to different stages of refactoring impact determination process**

### 3.2 Search strategy

We adopted the search strategy proposed by Martins and Gorschek [35] for the formulation and validation of search string required to search the relevant research articles. An overview of the search strategy employed in this work is provided online [36]. After string formation, the search for the identification of relevant papers was performed in two stages, firstly in August 2016 and later in January 2018. Each search stage followed two phases: (i) automatic search and (ii) manual search, along with reference (backward) snowballing.

### 3.2.1. Sources of information

In 2011, Zhang et al. [37] conducted a survey over a set of existing SLRs in the field of software engineering to determine the most frequently used electronic databases. In build-up to this work, we also performed a quick review of the digital libraries exploited in existing SLRs on software refactoring for extracting the most frequently utilized digital libraries. Based on this survey and experiences reported by Brereton et al. [38] and Dyba et al.[39], the following electronic databases were selected to conduct the search for relevant literature:

- IEEE Xplore
- ACM Digital Library
- ScienceDirect
- SpringerLink
- Scopus
- Wiley Online Library

Apart from these, other commonly used electronic databases are Kluwer Online, EI Compendex and Inspec. We excluded these databases from our search strategy because Kluwer Online has been merged with Springer and, EI Compendex and Inspec articles mostly overlap with Scopus articles [40].

*3.2.2. Search string formulation*

We followed the guidelines proposed by Kitchenham and Charters [31] to formulate our search string. The preliminary string formulation is performed through following steps:

- Firstly, the major search terms were drawn out from the research questions.
- Later, a list of abbreviations, synonyms and alternative spellings for the major search terms were obtained.
- Finally, major search terms were combined using boolean AND operator, and all alternative spellings, synonyms and abbreviations were joined using boolean OR operator to generate a preliminary search string.

The resulting preliminary string is: *(software **OR** program **OR** code **OR** class **OR** method **OR** package **OR** attribute **OR** function **OR** "object oriented" **OR** "object-oriented") **AND** (empirical\* **OR** "case study" **OR** "case studies" **OR** experiment\* **OR** survey) **AND** (refactor\* **OR** re-factor\*)*

**Table 2: Search strings defined for digital libraries with execution dates**

| Digital Library | Search String | Date |
|---|---|---|
| ScienceDirect | tak((software *OR* program *OR* code *OR* class *OR* method *OR* package *OR* attribute *OR* function *OR* {object oriented} *OR* {object-oriented}) *AND* (refactor\* *OR* {re-factoring} *OR* {re-factorization} *OR* {re-factorings} *OR* {re-factorizations} *OR* {re-factor} *OR* {re-factored} *OR* {re-factors})) | 7th Jan 2018 |
| IEEE Xplore | (software *OR* program *OR* code *OR* class *OR* method *OR* package *OR* attribute *OR* function *OR* "object oriented" *OR* "object-oriented") *AND* (refactor\* *OR* re-factor\*) | 7th Jan 2018 |
| SpringerLink | ((software *OR* program *OR* code *OR* class *OR* method *OR* package *OR* attribute *OR* function *OR* "object oriented" *OR* "object-oriented") *AND* (refactor\* *OR* "re-factor" *OR* "re-factoring" *OR* "re-factorings" *OR* "re-factored" *OR* "re-factorization" *OR* "re-factorizations" *OR* "re-factors")) | 7th Jan 2018 |
| Scopus | title-abs-key((software *OR* program *OR* code *OR* class *OR* method *OR* package *OR* attribute *OR* function *OR* {object oriented} *OR* {object-oriented}) *AND* (refactor\* *OR* {re-factoring} *OR* {re-factorization} *OR* {re-factorings} *OR* {re-factorizations} *OR* {re-factor} *OR* {re-factored} *OR* {re-factors})) | 8th Jan 2018 |
| ACM Digital Library | (Title:(software *OR* program *OR* code *OR* class *OR* method *OR* package *OR* attribute *OR* "object oriented" *OR* "object-oriented" *OR* function) *OR* Abstract: (software *OR* program *OR* code *OR* class *OR* method *OR* package *OR* attribute *OR* "object oriented" *OR* "object-oriented" *OR* function) *OR* Keywords: (software *OR* program *OR* code *OR* class *OR* method *OR* package *OR* attribute *OR* "object oriented" *OR* "object-oriented" *OR* function)) *AND* (Title:(refactor\* *OR* "re-factoring" *OR* "re-factorings" *OR* "re-factored" *OR* "re-factor" *OR* "re-factors" *OR* "re-factorization" *OR* "re-factorizations")) *OR* Abstract:(refactor\* *OR* "re-factoring" *OR* "re-factorings" *OR* "re-factored" *OR* "re-factor" *OR* "re-factors" *OR* "re-factorization" *OR* "re-factorizations")) *OR* Keywords:(refactor\* *OR* "re-factoring" *OR* "re-factorings" *OR* "re-factored" *OR* "re-factor" *OR* "re-factors" *OR* "re-factorization" *OR* "re-factorizations"))) | 8th Jan 2018 |
| Wiley Online Library | ((software *OR* program *OR* code *OR* class *OR* method *OR* package *OR* attribute *OR* function *OR* "object oriented" *OR* "object-oriented") in Article Titles *OR* (software *OR* program *OR* code *OR* class *OR* method *OR* package *OR* attribute *OR* function *OR* "object oriented" *OR* "object-oriented") in Abstract *OR* (software *OR* program *OR* code *OR* class *OR* method *OR* package *OR* attribute *OR* function *OR* "object oriented" *OR* "object-oriented") in Keywords) *AND* ((refactor\* *OR* re-factor\*) in FullText) | 8th Jan 2018 |

In order to validate the search terms of the initial search string, we manually selected a set of 20 articles from our database containing pre-collected most relevant articles (related to refactoring and software quality). The constructed search string needs to search against the titles, abstracts and keywords of the articles included in each of the selected electronic databases. Therefore, we checked the occurrences of any of the search string terms against the title, abstract and keywords of the 20 nominated articles. We believe such a validation would assure us of the complete coverage of relevant literature published in selected digital libraries. However, only 11 out of all 20 nominated articles were captured by our initial search string, which influenced us to optimize our search string to a more generic string so as to enhance the article coverage. All the data related to the validation of our search string terms is available online [36]. The resulting general search string obtained after extensive experimentation is as follows:

*(software **OR** program **OR** code **OR** class **OR** method **OR** package **OR** attribute **OR** function **OR** "object oriented" **OR** "object-oriented") **AND** (refactor\* **OR** re-factor\*)*

Further, as different digital libraries follow different search query formation/syntax rules, we defined dedicated search strings for individual digital libraries (as provided in Table 2) considering the recommendations and cautions provided by Singh and Singh [41]. Before moving to the data collection process, we also performed a pilot study using the formulated search strings in selected digital libraries and compared the result set of trial search with 20 manually generated reference articles. The search string captured 18 out of 20 reference articles, which validates our search string. The remaining two reference articles [42-43] could not be identified due to their non-existence in any of the selected digital libraries.

### 3.2.3. Automatic search

In this phase, we executed our search strings (shown in Table 2) restricted to the title, abstract and keywords of the articles published in six electronic sources namely IEEE Xplore, ACM Digital Library, Science Direct, SpringerLink, Scopus and Wiley Online Library. The first round (Phase 1) of automatic search was completed in August 2016 returning an initial set of 12,996 search results. To update this SMS further till December 2017, another round (Phase 2) of search was conducted in January 2018 which rendered additional 287 search results. Hence, both the phases resulted in a total of 13,283 search results, which were managed and organised using Zotero[1]. We faced some challenges with SpringerLink, as it does not provide the option of restricting the search to title, abstract and keywords. We were hence forced to conduct a full text (including references) level search, which resulted in 6,698 results.

### 3.2.4. Manual search

To ensure the completeness of our article list, we also performed a manual search. Manual search was conducted in renowned conference proceedings and journal issues published between January 2005 and December 2017. The selected set of conference proceedings and journals for the manual search are listed in Table 3. These sources were selected because they have contributed a large number of research articles in the area of software refactoring. The title of each conference or journal article was independently reviewed by both the authors of this paper. Manual search resulted in 174 articles (161 and 13 research articles from both Phase 1 and Phase 2, respectively), which were recorded in MS Excel spreadsheets after undergoing a scrutiny against our inclusion and exclusion criteria (Section 3.3.1).

Table 3: Selected journals and conference proceedings for manual search

| Source Type | Source Name | Acronym |
|---|---|---|
| Conference proceedings | • International Conference on Software Maintenance and Evolution (Formerly known as International Conference on Software Maintenance) | ICSME (ICSM) |
| | • International Conference on Software Engineering | ICSE |
| | • European Conference on Software Maintenance and Reengineering | CSMR* |
| | • Working Conference on Reverse Engineering | WCRE* |
| | • Automated Software Engineering | ASE |
| Journals | • IEEE Transactions on Software Engineering | TSE |
| | • ACM Transaction on Software Engineering and Methodology | TOSEM |
| | • Journal of Systems and Software | JSS |
| | • Information and Software Technology | IST |
| | • Empirical Software Engineering | EMSE |

*CSMR and WCRE were combined in International Conference on Software Analysis, Evolution, and Reengineering (SANER) from year 2014. Hence, we explored SANER proceedings from 2014 onwards.

Thereafter, the results of automatic and manual search phases were combined, and also compared to eliminate all the duplicate results. These results were further filtered independently by both the authors on the basis of title, abstract and full text, which left us with a set of 129 (100 and 29 research articles from both Phase 1 and Phase 2, respectively) relevant articles related to our review scope. To evaluate the agreement between both the authors, Cohen's Kappa statistics [44] was employed, as suggested by Kitchenham and Charters [31]. In cases of disagreement, mutual discussions between the two authors were conducted to resolve the differences and converge to a common solution. The complete article screening process is reported in Section 3.3.2.

Furthermore, out of 174 relevant research articles obtained after manual search, only one PS [45] was included into the final pool of 142 PSs after the article screening process (Section 3.3.2). As our automatic search is restricted to title, abstract and keywords only, it [45] did not contain the term 'refactor' within the same and hence did not satisfy our search criteria.

---

[1] http://www.zotero.org

*3.2.5. Reference checking*

Both authors independently performed the reference snowballing by manually checking the reference list of the 129 relevant articles obtained following the refinement of automatic and manual search results. Furthermore, the references of previously published SLRs [18-21, 23] in the area of software refactoring were also checked to mitigate the risk of missing any relevant articles. The results obtained after reference snowballing were matched and discussed between the authors to obtain a coherent list of research articles. This additional phase helped us in reducing the risk of missing any relevant literature.

**3.3.** *Study selection process*

This section describes the study selection criteria followed to filter the primary studies from the potentially relevant articles.

*3.3.1. Inclusion and exclusion criteria*

The inclusion and exclusion criteria applied to screen out the articles that are not significant to the defined SMS research questions, are listed in Table 4. Before executing study selection process, both the authors independently verified the inclusion and exclusion criteria on a set of 90 articles randomly selected from automatic search results obtained after Phase 1. The agreement between both the authors was evaluated using Cohen Kappa statistic [44]. The first attempt rendered Kappa=0.57 which was classified into *moderate* category according to Landis and Koch [46]. This *moderate* agreement between both the authors may be attributed to difference in opinions concerning the meaning of inclusion and exclusion criteria. Hence, to develop a common understanding about content-related criteria and resolve the disagreements, a set of meetings and diligent discussions were conducted between the authors, and the inclusion criteria was refined accordingly. Later, we again applied Cohen Kappa statistic [44] on another set of randomly selected 90 articles and obtained Kappa=0.77 (*substantial*). This substantial agreement indicates that the inclusion and exclusion criteria is clear enough to both the authors, and can now be applied more efficiently and reliably to the article screening process performed in Section 3.3.2. We did not apply Cohen Kappa coefficient before screening research articles in Phase 2, as we already obtained a refined criteria after Phase 1.

**Table 4. Inclusion and exclusion criteria**

| Inclusion Criteria | Exclusion Criteria |
|---|---|
| • Research articles published in peer-reviewed symposiums, workshops, magazines, journals, transactions or conference proceedings before January 2018.<br>• Studies written in English language.<br>• Studies that provide research methodology or tool for determining the refactoring impact on object-oriented software quality.<br>• Studies specific to performing the refactoring activities that eliminate code bad smells. | • Poster sessions, prefaces, article summaries, position papers, book reviews, discussion, editorials, readers' letters, panels, and conference companions and summaries of tutorials, workshop and symposiums.<br>• Quantitative or qualitative studies reporting results without providing supporting evidence.<br>• Studies with the non-availability of full text.<br>• Duplicate research articles, e.g.<br>  ➢ Studies with their extended versions published in different publication venues,<br>  ➢ Studies found identical in different digital libraries.<br>• Studies concerned with the refactoring of software artifacts other than source code, e.g.<br>  ➢ Requirement specifications,<br>  ➢ Design (UML models),<br>  ➢ Test cases,<br>  ➢ Database schemas, etc.<br>• Studies specific to performing the refactoring activities not related to code bad smells, e.g.<br>  ➢ Cross paradigm refactoring (object oriented source code to aspect-oriented or service-oriented code),<br>  ➢ Serial code to parallel code, etc.<br>• Studies that consider non object-oriented software, e.g.<br>  ➢ Functional programming,<br>  ➢ Service-oriented code,<br>  ➢ Aspect-oriented code, etc. |

Regarding the last inclusion criterion, it is to be noted here that the main motivation to perform refactoring activities in the selected PSs, is to remove code smells. This motivation was not clearly stated in many of the PSs. In such cases, the studies which applied refactoring activities proposed by Fowler [8] to investigate their impact on software quality are included. This approach was followed because it is explicitly indicated in Fowler's book that the listed refactoring activities target code smells. Further, we are not interested in the papers that report the impact of change in coupling/cohesion on refactoring decisions [47-48]. Rather, our focus is on determining the impact of refactoring activities on internal/external software quality attributes. Furthermore, the studies reporting tools for code smell detection or refactoring opportunity identification, which do not assess or predict the impact of refactoring activities on software quality are excluded.

*3.3.2. Article screening*

The article screening process used to select the primary studies consists of seven stages. The search for the relevant literature was firstly conducted in August 2016, and the results were later updated in January 2018 for covering the literature published till December 2017. Figure 4 shows the article screening process along with the article count retrieved at the end of each stage for both the aforementioned phases. At Stage 1, we conducted the automatic search in six electronic databases together with the manual search in selected journals and conference proceedings. The search results captured using automatic search were extracted using Zotero[1], and later exported to an MS Excel spreadsheet. In Stage 2, data cleaning of automatic search results was performed to filter out the irrelevant entries like bibliography, workshop/symposium summaries, etc.

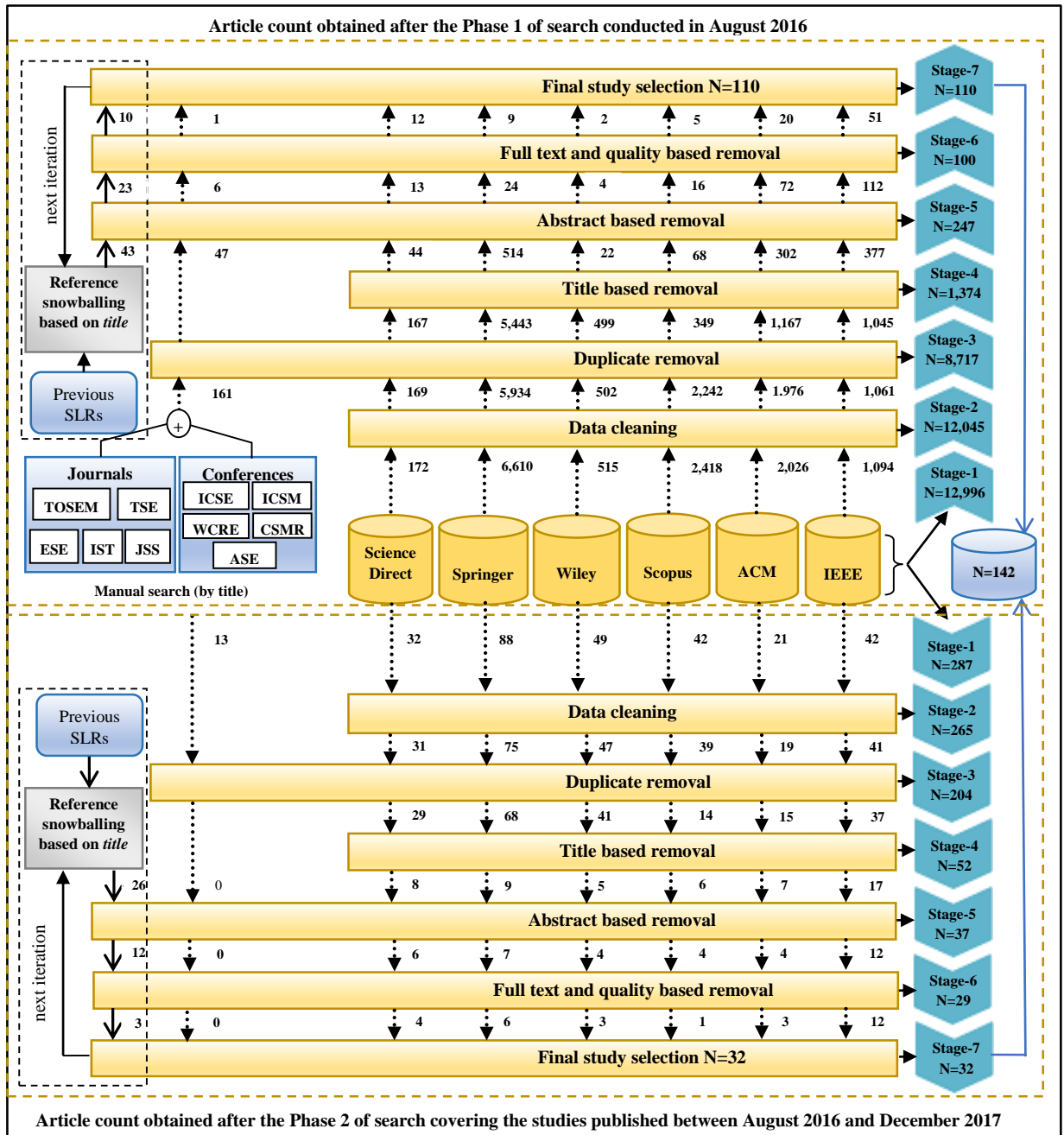

**Figure 4: Stages of article selection process**

In Stage 3, we firstly merged the results of both automatic and manual searches in Excel spreadsheet. Later, we manually eliminated the duplicate papers from this spreadsheet after sorting them alphabetically by article titles. In subsequent stages, both the authors independently filtered the remaining search results on the basis of title, abstract and full text, following the inclusion/exclusion criteria defined in Section 3.3.1. At each stage, we used

Cohen's Kappa statistics [44] to evaluate the homogeneousness between both the authors as suggested by Kitchenham and Charters [31]. The details regarding the level of agreement between both the authors at each stage of article screening process are shown in Table 5.

In Stage 4, we discarded the articles based on their titles. The statistical analysis resulted in a kappa value of 0.78 and 0.73 for Phase 1 and Phase 2, respectively which signifies a *substantial* agreement in both cases, according to the categories proposed by Landis and Koch [46]. In case of disagreement, we decided to include the research articles to next stage in order to abate the risk of eliminating the relevant research articles as recommended by Petersen and Ali [49]. This process selected 1374 (958+416) and 52 (33+19) research articles from Phase 1 and Phase 2, respectively for further screening in Stage 5.

In Stage 5, the shortlisted articles were further screened by both the authors independently on the basis of their abstracts. The corresponding strength of agreement between both the authors was found to be *substantial* for both the phases. In case of uncertainty, the research articles selected by either of the two authors were provided as an input to Stage 6. In Stage 6, both the authors individually performed the inclusion and exclusion process by reading the full text of 247 (169 +78) and 37 (30 + 7) articles obtained from Phase 1 and Phase 2, respectively of Stage 5, to further exclude the irrelevant articles. To deal with lack of congruence between both the authors, we discussed regarding the inclusion of each of the 20 (17 in Phase 1 and 3 in Phase 2) studies and finally decided to include 8 (6 in Phase 1 and 2 in Phase 2) studies. Hence, this process included 114 (108+6) and 31 (29+2) articles from Phase 1 and Phase 2, respectively, after full text based removal.

In addition, these articles were further assessed for quality by utilizing the quality assessment checklist described in Section 3.3.3 and the obtained values of Cohen's Kappa coefficient at this stage, are enlisted in Table 5. We conducted a set of meetings and discussions to resolve the disagreement on 6 (5 in Phase 1 and 1 in Phase 2) articles and then agreed to include 3 (2 in Phase 1 and 1 in Phase 2) articles. This resulted in a total of 100(98+2) and 29 (28+1) articles from Phase 1 and Phase 2, respectively into the final catalogue of selected PSs. In Stage 7, we checked the references of shortlisted articles and previously published relevant SLRs [18-21, 23] based on their titles. These *reference snowballing* results were further refined on the basis of abstract and full text. Finally, the selected articles obtained after Stage 6 (including reference snowballing) were included in the final list of 142 primary studies.

Table 5: Agreement between both the authors during article screening process

| Article Screening Stage | Phase | # Input Studies | Agreement for Inclusion | Agreement for Exclusion | Disagreement | Kappa Value | Level of Agreement |
|---|---|---|---|---|---|---|---|
| Stage 4- Title based removal | Phase 1 | 8717 | 958 | 7343 | 416 | 0.78 | Substantial |
| | Phase 2 | 204 | 33 | 152 | 19 | 0.73 | Substantial |
| Stage 5- Abstract based removal | Phase 1 | 1374 | 169 | 1127 | 78 | 0.77 | Substantial |
| | Phase 2 | 52 | 30 | 15 | 7 | 0.71 | Substantial |
| Stage 6- Full text based removal | Phase 1 | 247 | 108 | 122 | 17 | 0.86 | Perfect |
| | Phase 2 | 37 | 29 | 5 | 3 | 0.72 | Substantial |
| Stage 6- quality based removal | Phase 1 | 114 | 98 | 11 | 5 | 0.81 | Perfect |
| | Phase 2 | 31 | 28 | 2 | 1 | 0.79 | Substantial |

### 3.3.3. *Study quality assessment*

While conducting an SLR, it is important to analyse the quality of studies in order to weigh the significance of each study when the results are being synthesized. This assists in selecting high quality studies for deriving reliable results and conclusions [31]. Study quality assessment is not a mandatory practice in case of a SMS. However, based on some recent SMS [50-53], we decided to include a quality assessment step into our mapping protocol. Based on Kitchenham and Charters guidelines [31], we firstly constructed a quality assessment checklist comprising of ten questions that cover several study aspects including design, conduct, data analysis and conclusion. Later, we read the full text of each study to answer the quality checklist questions. For each question, each study is evaluated by both the authors iteratively (i.e. the first author assessed the quality of each study and the results were later verified by the second author) on a three-point scale with '*Yes*', '*Partially*' or '*No*' grade. Each question was further quantified by designating a numeric value of 0, 0.5 and 1 to '*Yes*', '*Partially*' and '*No*' grade, respectively. The final quality score of a study was later obtained by adding up the scores of all quality assessment questions corresponding to that study, and the studies that scored more than 4.5 qualified to be included in the final list. The complete information regarding quality assessment grades assigned to each PS is provided online [36]. Furthermore, the procedure used to avoid the inconsistency of quality assessment process is detailed in Section 3.3.2. Some of the questions were not applicable to a few primary studies, hence we used a '*NA*' value on the point scale for such cases. The results of this quality assessment step are reported in Table 6.

The results in Table 6 show that most of the quality assessment questions (QA1, QA3, QA6, QA9 and QA10) received positive answers. For QA1, we reviewed the abstract and introduction sections of each study to check whether the research aims are clearly stated or not. The 142 primary studies either adequately (89.4%) or partially (10.6%) described the research objectives. For QA2, we assessed whether the authors of the study clearly described the research methodology used for determining the impact of refactoring activities on software quality. For this, we looked at the study design and methodology sections, and found that 67.6% of the studies adequately explained the research methodology.

The complete structure of the paper was reviewed in order to answer QA3-QA7. About 91.5% of studies have clearly stated the software quality attributes or measures on which the impact of refactoring activities is measured (QA3) and 69.7% of studies defined these software quality attributes or measures appropriately (QA4). Regarding QA5 and QA6, we noticed that a considerable number of studies (76% and 92.3%, respectively) explicitly defined the datasets in terms of size and programming language used. Concerning QA7, we found that very few (only 18.3%) PSs measured the statistical significance of the outcomes obtained regarding the effect of refactoring on software quality. For QA8, we reviewed the discussion, limitations and threats to validity sections of the papers, and observed that the threats to validity are scarcely discussed. We answered this question with *'Yes'* for the studies where threats to validity are explicitly discussed; and marked *'Partially'* for the studies that mentioned threats to validity without proper explanation. For QA9 and QA10, we looked into the main findings, discussion, conclusions and future work sections of the papers to determine whether all study questions are answered and the research findings are properly documented. For most of the studies, we received positive answers (99.3% and 97.2%, respectively) for these questions.

**Table 6: Quality assessment results**

| ID | Question | Percentage of PSs | | | |
|---|---|---|---|---|---|
| | | Yes | Partially | No | NA |
| *Research design & conduct* | | | | | |
| QA1 | Are the research aims of the study clearly stated? | 89.4% | 10.6% | 0% | 0% |
| QA2 | Is the research methodology for determining the impact clearly described? | 67.6% | 28.2% | 4.2% | 0% |
| QA3 | Are the internal/external software quality attributes or measures on which the impact is measured clearly stated? | 91.5% | 8.5% | 0% | 0% |
| QA4 | Are the definitions of the quality attributes or measures provided? | 69.7% | 20.4% | 9.9% | 0% |
| *Data analysis* | | | | | |
| QA5 | Is the size of datasets adequately described? | 76% | 8.5% | 14.8% | 0.7% |
| QA6 | Is the programming language of the datasets stated? | 92.3% | 3.5% | 3.5% | 0.7% |
| QA7 | Are results statistically significant? | 18.3% | 0% | 81.7% | 0% |
| *Conclusion* | | | | | |
| QA8 | Are the validity threats/ limitations discussed? | 54.9% | 9.9% | 35.2% | 0% |
| QA9 | Are all study questions answered? | 99.3% | 0.7% | 0% | 0% |
| QA10 | Do empirical data and results support the conclusions? | 97.2% | 2.8% | 0% | 0% |

### 3.3.4. Mapping primary studies

We assigned unique identifiers ranging from [S1] to [S142] to the final set of selected primary studies. The information regarding the mapping of the PSs to their respective identifiers (IDs) is provided online [36]. Henceforward in this SMS, each primary study will be referred using its unique ID ([S1], [S2], etc.). It should be noted that among the studies passing Stage 6 of the article screening process, we found some duplicate articles published by the same authors in different venues as an extension to their prior works. For example, an extension of an article published in some conference or workshop proceedings found in a journal publication by performing additional case studies or experiments while maintaining the original context (for instance, two contributions by Cinnéide et al. [54-55] are merged into one PS [S73])**.** Based on Kitchenham and Charters guidelines [31], two or more articles published by the same authors focusing on similar approach or similar context are considered as one primary study. This is attained by allocating the union of their attribute subsets to the most recent publication and discarding the other publications at hand [56]. This approach resulted in a manageable number of primary studies without the loss of any information. The information regarding the articles merged as one primary study is available online [36].

### 3.4. *Data extraction and synthesis*

We extracted the relevant data from the final set of 142 articles in order to answer the research questions defined in Section 3.1. The information extracted after reading each article in depth was recorded in a data extraction form. Following items were extracted from each PS:

- Full reference information of primary study, including title, author, publication title and publication year
- Internal/external quality attributes on which the impact of refactoring activities is determined (Section 3.5.6)
- Research contribution method (Section 3.5.1)
- Study context/setting (Section 3.5.2)
- Software quality measures used as surrogates to determine the effect of refactoring activities on software quality (Section 3.5.4)
- Refactoring activities (Section 3.5.3)
- Statistical techniques (Section 3.5.5)
- Dataset characteristics like name, size, type and programming language (Section 3.5.7)
- Empirical results concerning the influence of refactoring activities on software quality (Section 4.2.5)

The details of data extraction form given in an online appendix [36] further assisted us in classifying the PSs into different categories as discussed in Section 3.5. After data extraction, we observed that only 18.3% of the selected PSs applied statistical techniques to determine the statistical significance of results regarding the effect of refactoring activities on software quality. In addition, due to the diversity of software quality measures and refactoring activities, meta-analysis or combining p-values becomes an inapplicable approach while merging the outcomes related to impact of refactoring on software quality. Therefore, we opted for *vote-counting* approach because it was the only possible choice in this situation, as also mentioned by Dallal and Abdin [21]. Vote-counting approach is widely adopted in many similar SLRs [57-58] and empirical studies [59-61] conducted in the area of software engineering and it does not take software quality measures or dataset sizes into consideration while merging the results [21].

### 3.5. *Study classification*

In this SMS, we followed a 4-phase approach to classify the selected PSs into seven facets namely *research contribution method, study context, refactoring activities, software quality measures, investigation approach, focus,* and *dataset.* Firstly, both the authors independently started jotting down all the possible categories to which the selected PSs may fall, by considering both the research questions identified in Section 3.1 and the taxonomies provided by previous relevant works [19, 62-63]. Following this approach, the common categories among both the authors were initially chosen for further investigation. Later, both the authors resolved the cases of discrepancies through diligent discussions and finally reached a consensus to include aforementioned seven categories. Secondly, the initial set of possible values for each category was established based on preliminary analysis performed on previous SLRs [19, 25, 28, 62-63]. Thirdly, we extracted the data by reading the full text of each selected PS to determine the exact set of values that each facet can hold as all the selected PSs cannot fit into predetermined set of values. Based on Staples and Niazi [64] guidelines, the first author performed the data extraction of each study considering the above mentioned seven categories and their predefined values and the results were later checked by the second author. During data extraction, new value of each category was added to the catalogue of possible values only if it was not present in the initial list. Furthermore, in the event of lack of congruence between the authors, a set of discussions were conducted to ensure mutual conformity. After data extraction, we analysed the data value corresponding to each category and a particular value which was available in the initial list but not in our data extraction form, was eventually discarded. Finally, we grouped the selected PSs based on the finalized consistent list of facets and their sub-facets. The study classification scheme is depicted in Figure 5-6.

### 3.5.1. *Research contribution method*

We adopted the categorization used by Wangberg [19] and Bissi et al. [62] to divide the studies according to applied research method. The *research contribution* facet contains two broad subcategories: *empirical research* and *tool. Empirical research* derives knowledge from empirical evidences rather than from theory or belief. This sub-facet involves *case study, experiment, simulation*, or *survey* conducted to validate the effect of refactoring activities on software quality. *Tool* sub-facet refers to a software tool provided to assess/predict the impact of refactoring activities on software quality. It is evident from Figure 5a that *case study* is the most commonly utilized empirical research method (81%) followed by the *experiments* (12.7%). We also found that nearly 4.9% of the PSs have used more than one empirical research method to validate the influence of refactoring activities on software quality. Finally, *tools* have been proposed by 7.7% of the selected PSs.

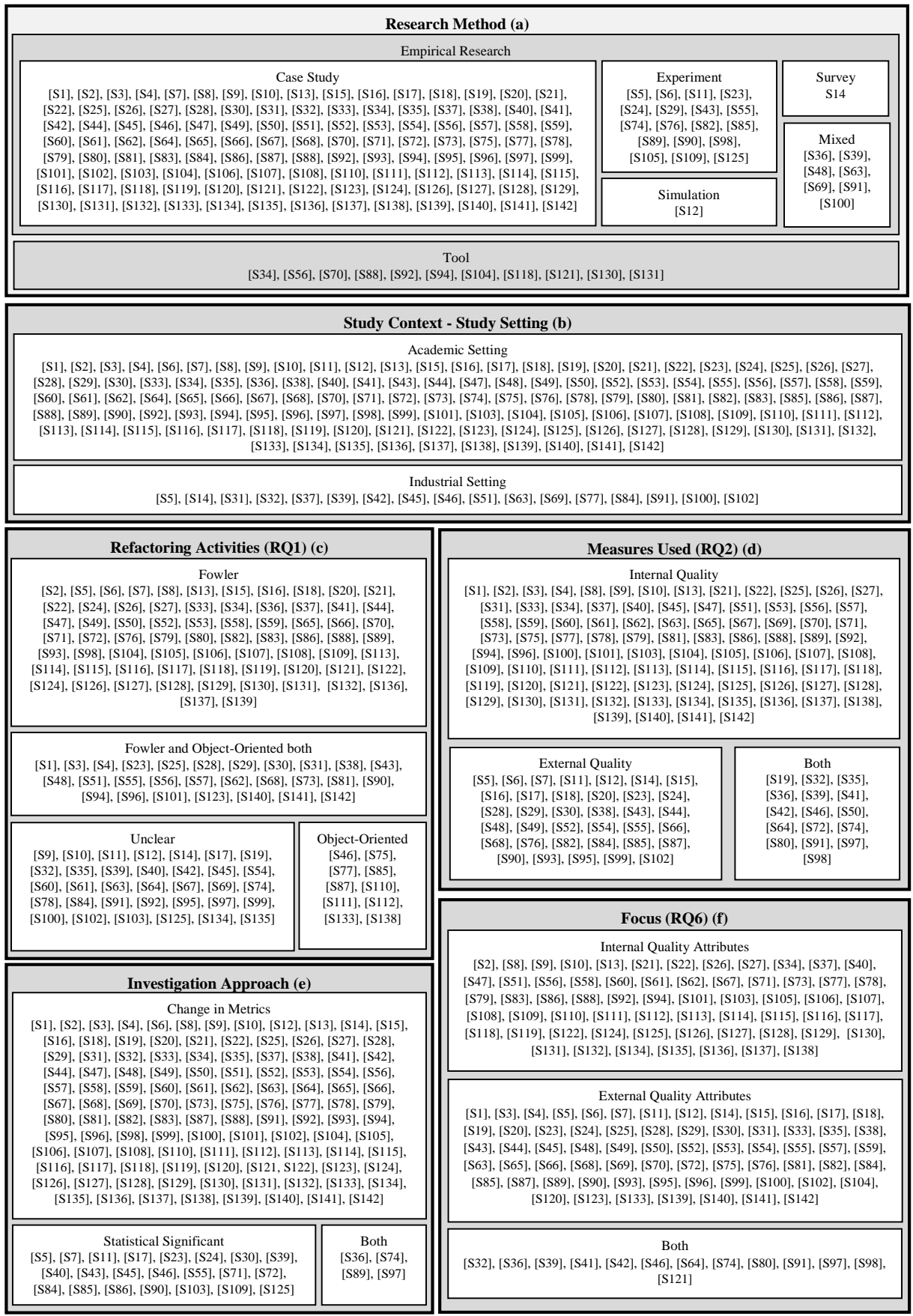

**Figure 5.** Study classification scheme facets along with PSs mapping based - a) research contribution method, b) study context: study setting, c) refactoring activities, d) measures used, e) investigation approach, f) focus

*3.5.2. Study context*

We divided the *study context* facet into three main sub-facets namely *study setting, author setting type* and a*uthor setting location. Study setting* specifies the context in which a study is conducted in order to assess the external validity of findings. Based on Bissi et al. [62] suggestions, PSs were divided into two categories namely *academic setting* and *industry setting*. Academic studies consider the research performed by the university professionals or research organisations in scholastic world, whereas industrial studies consider the research conducted by software practitioners in industrial environment. It can be observed from Figure 5b that most of the studies (88%) were performed in academic environments, and only 12% studies were conducted in commercial organisations. Furthermore, a*uthor setting type* and a*uthor setting location* refers to the environment and geographical areas, respectively with which the authors of selected PSs are associated. Due to space limitations, the classification of selected PSs against these facets is provided as a part of the online supplementary material [36].

*3.5.3. Refactoring activities*

The term 'refactoring activities' in this text refers to refactoring techniques or refactoring operations like *Extract Method, Move Method*, etc. [25]. Only the refactoring activities used to remove code smells are considered in this SMS. After screening all PSs, we noticed that 24% PSs did not clearly mention the considered refactoring activities. Among the rest (76%) of the PSs, a total of 154 distinct refactoring activities are used, out of which 70 refactoring activities are proposed by Fowler [8]. The considered refactoring activities are listed in detail in Section 4.2.1.

*3.5.4. Software quality measures*

Quality measures can be categorized as *internal* or *external* measures. *Internal quality measures* can be objectively measured from the source code of the software. For instance, LCOM is an internal measure of software product. *External quality measures* can be assessed from the external behaviour of the software, and are described stochastically due to their dependence on human and environmental aspects [65]. For example, execution time is an external measure of software performance. The detail of the software quality measures considered by selected PSs to determine the impact of refactoring activities on software quality is provided later in Section 4.2.2.

*3.5.5. Investigation Approach*

In this facet, we classified the selected PSs into *statistical significant* and *change in metrics* sub-facets. The studies which applied statistical techniques to explore the effect of refactoring on software quality are placed in the former category. On the contrary, the studies which reported the impact of refactoring on software quality simply in terms of change in the values of software quality measures without reporting the application of any statistical methods are placed in the latter category. The complete information regarding the applied statistical approaches is given in Section 4.2.5.

*3.5.6. Focus*

The *focus* of a study is a particular type of quality attribute that the study has targeted. This facet helped us in categorizing the studies more specifically according to internal and external quality attributes on which the impact of refactoring activities is measured. In this work, we followed the notion of *software quality attributes* as explained by Fenton and Pfleeger [65] and Morasca [66], along with the previous highly-cited systematic reviews [21, 57]. Among 142 PSs, 62 PSs reported the impact of refactoring activities only on internal quality attributes and 13 PSs determined the impact of refactoring activities on both internal as well as external quality attributes. The rest 67 PSs measured the effect of refactoring activities on external quality attributes by using either internal quality measures as surrogates or external quality measures.

*3.5.7. Dataset*

The term *dataset* means a single subject software system. This facet is further categorized into three sub-facets namely *dataset size, type* and *domain*. We followed the guidelines provided by Radjenovic et al. [63] to classify the studies into *small, medium* and *large* sub-dimensions based on their dataset sizes in terms of number of classes (or files) and number of Lines of Code (LOC) by following Equation 1.

$$Dataset\ Size = \begin{cases} Small & KLOC\ (thousand\ LOC) < 50\ OR\ Classes < 200 \\ Medium & 50 \leq KLOC \leq 250\ OR\ 200 \leq Classes \leq 1000 \\ Large & KLOC > 250\ OR\ Classes > 1000 \end{cases} \quad (1)$$

Further each PS just needed to satisfy one of the two criteria (KLOC/number of classes) to be placed in a higher size category. For instance, a PS that used a subject system (for e.g. JMeter in [S52]) with 40 KLOC and 400 classes, is categorized as *medium*. Based on the categorization of Al Dallal [25], we also divided the datasets into *academic, student, commercial* and *open source* project types. To identify the attributes of domain sub-facet, we screened the data of all PSs from the data extraction form. This process resulted in mainly five programming language attributes: *Java, C#, C++, Python* and *Swift*. Furthermore, the PSs which failed to report the dataset attributes like size, language or type are placed against *not informed* (NI) category. The detail of the explored datasets is presented in Section 4.2.4.

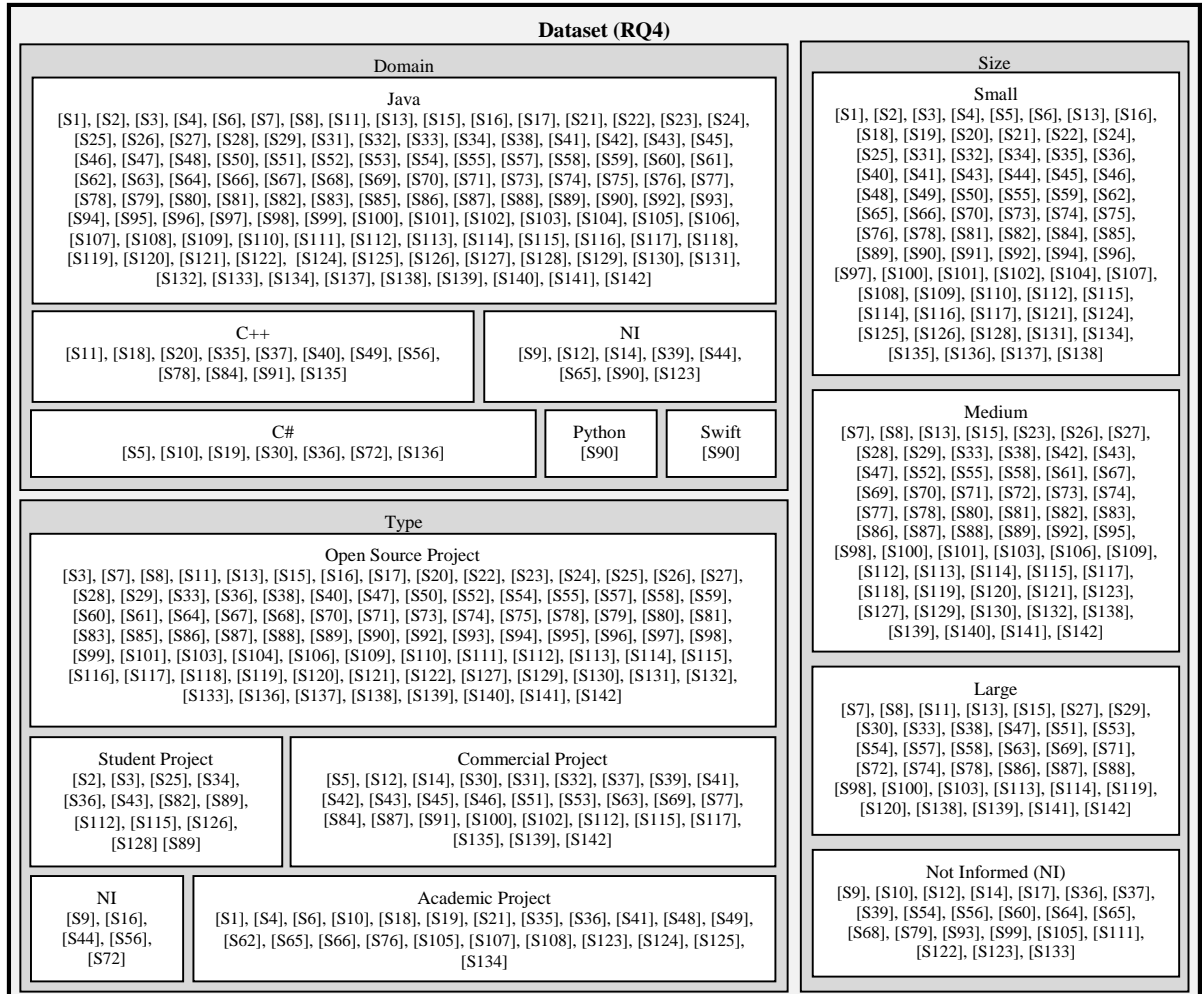

**Figure 6. Study classification of dataset facet along with PSs mapping**

## 4. RESULTS

The main findings derived from the information extracted from the selected set of PSs are reported in this section. Firstly, the distribution of the selected 142 PSs across publication year, type and venue supported by appropriate data representations in the form of tables and graphs is presented. In the later subsections, the extracted data is analysed to provide detailed answers to the research questions defined in Section 3.1.

*4.1. Overview of PSs*

The distribution of the selected PSs mapped to their respective publication years is shown in Figure 7. Figure 7 indicates that there were no relevant publications during the first 8 years after the term refactoring was formulated. This is primarily either because the field of refactoring became substantial in 1999 after the publication of Fowler's refactoring book, or the studies published during that period did not pass our inclusion criteria.

Furthermore, we observed that the number of PSs between the years 2000 and 2004 varies from one to three publications. But the noticeable increase in the number of PSs thereafter except for the years 2007 and 2013 may be attributed to any of the following reasons.

- the interest in the area of assessing the impact of refactoring activities on software quality has been gaining new grounds in the last few years.
- amelioration in the number of researchers working in the area of software engineering over the passage of time, having higher publication pressure.

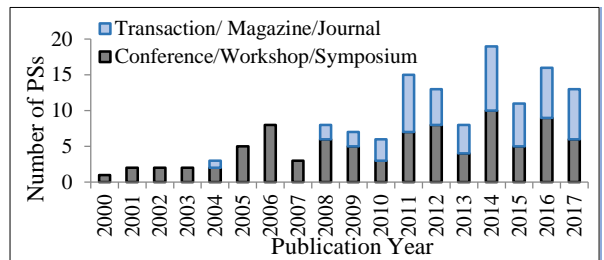

Figure 7: Distribution of PSs over publication years

Table 7: Distribution of PSs over publication type

| Publication Type | No. of PSs | Percentage |
|---|---|---|
| Conference | 68 | 48% |
| Journal | 43 | 30.3% |
| Workshop | 10 | 7% |
| Symposium | 10 | 7% |
| Transaction | 8 | 5.6% |
| Magazine | 3 | 2.1% |

The distribution of the selected PSs across various publication types is represented in Table 7. A relatively high number of PSs (48%) were published in conference proceedings. The first journal primary study was published in 2004, and over half of the journal PSs (53.5%) were published between the years 2014 and 2017. Each publication type is represented by at least three articles in the final set of 142 PSs. The publication venues contributing more than two PSs are enlisted in Table 8. Overall, our 142 PSs are scattered across 85 different publication sources which indicates that the research related to measuring the effect of refactoring activities on software quality has gained extensive attention in the entire software engineering community. Also, *Information and Software Technology* followed by *IEEE International Conference on Software Maintenance and Evolution* (formerly known as *IEEE International Conference on Software Maintenance*) are found to be the most prominent venues to publish the research concerning the impact of refactoring on software quality. The venues which contributed two or less publications are reported as '*Others*' in Table 8.

Table 8 - Distribution of PSs over publication venue

| Type | Publication Venue Title | PSs |
|---|---|---|
| Transaction/ Journal/ Magazine | • Information and Software Technology | 8 |
| | • Journal of Systems and Software | 6 |
| | • IEEE Transactions on Software Engineering | 5 |
| | • Empirical Software Engineering | 5 |
| | • Journal of Software: Evolution and Process | 3 |
| | Others | 27 |
| Conference/ Symposium/ Workshop | • IEEE International Conference on Software Maintenance and Evolution | 7 |
| | • European Conference on Software Maintenance and Reengineering | 6 |
| | • Asia-Pacific Software Engineering Conference | 4 |
| | • Working Conference on Reverse Engineering | 3 |
| | • International Working Conference on Source Code Analysis and Manipulation | 3 |
| | • International Conference on Product-Focused Software Process Improvement | 3 |
| | • International Conference on Quality of Information and Communications Technology | 3 |
| | • International Conference on Software Analysis, Evolution and Reengineering | 3 |
| | Others | 56 |

*4.2. Answer to research questions*

All 142 shortlisted PSs were thoroughly reviewed in order to extract the relevant data required to answer the set of research questions defined in Section 3.1. This subsection presents the answers to our formulated research questions.

*4.2.1. RQ1:* Which refactoring activities and underlying code smells were investigated?

The selected PSs used several approaches to retrieve the information regarding refactoring activities for measuring their impact on software quality. The first approach is to apply specific refactoring activities to software systems either manually or by using automated tools. Another approach is to mine the applied refactoring activities by: a) searching for the keywords like 'refactor' from commit logs, b) analysing the code history, c) observing the software developers, or d) using a tool like Ref-Finder [67] from version archives of software systems. After data analysis, we found that 111 PSs (78.2%) opted for first approach, whereas 32 PSs (22.5%) used the latter approach. Furthermore, the studies which extracted the refactoring activities by using the second approach, did not mention the names of code smells targeted by software practitioners. This may be attributed to following reasons: a) software developers commit the performed changes by simply mentioning the name of refactoring activity rather than also giving the information about removed code smells, and b) refactoring detection tools like Ref-Finder [67] only provide detail about refactoring activities.

| God Class | Duplicate Code | Functional Decomposition | Shortgun Surgery | Data Class | Complex Conditional | Internal or private Getter and Setter | Divergent Change |
|---|---|---|---|---|---|---|---|
| [S24], [S26], [S27], [S40], [S50], [S59], [S63], [S65], [S74], [S80], [S90], [S100], [S105], [S109], [S113], [S114], [S116], [S118], [S127], [S128], [S137], [S139], [S140], [S141], [S142] | [S27], [S41], [S43], [S59], [S63], [S65], [S82], [S83], [S100], [S111], [S121], [S123], [S124], [S131], [S134], [S135] | [S139], [S140], [S142] | [S26], [S35], [S100], [S105], [S141], [S142] | [S27], [S53], [S89], [S139], [S140], [S141], [S142] | [S18], [S117], [S122] | [S85], [S90] | [S35], [S53], [S128] |
| | Feature Envy | Long Method | Long Parameter List | Spaghetti Code | Lazy Class | Middle Man | Data Clumps | Hash Map Usage |
| | [S26], [S41], [S43], [S65], [S70], [S80], [S89], [S115], [S120], [S125], [S126], [S131], [S130], [S136], [S139], [S141], [S142] | [S26], [S41], [S63], [S65], [S80], [S100], [S105], [S125], [S131], [S132] | [S63], [S65], [S90], [S100], [S125], [S141] | [S51], [S74], [S139], [S140], [S141], [S142] | [S90], [S105], [S141] | [S53], [S88] | [S63], [S70], [S100] | [S85], [S90] |

**Figure 8: Distribution of code smells**

Furthermore, out of the 111 PSs which applied refactoring activities either manually or automatically to a dataset, 55 PSs did not highlight the names of targeted code smells. Among the rest 56 PSs, 31 code smells were considered. The code smells which were targeted by more than one PS are shown in Figure 8 and the complete information about them is provided online [36]. It is to be noted that as only 39% PSs provided the information about code smells, we could not conclude anything substantial from such a few number of PSs. Therefore, we will not discuss code smells henceforward in this paper.

During data extraction, we found that the selected 142 PSs either clearly or partially specified, or did not specify the names of considered refactoring activities. Accordingly, we classified these PSs into three main categories namely *specific, partially specific,* and *general*. The studies which have clearly mentioned the exact refactoring activities considered to determine the impact of refactoring activities on software quality fall under the *specific* category. The PSs in which the authors provided incomplete information about the considered refactoring activities are placed under *partially specific* category. Finally, the studies in which the authors have not mentioned particular refactoring activities exercised, are grouped in *general* category. Figure 9a depicts that about 70%, 6% and 24% of PSs belong to *specific, partially specific,* and *general* categories, respectively. Moreover, the 108 PSs under *specific* and *partially specific* categories are classified into *Fowler* (proposed by Fowler [8]), *object-oriented* (defined to remove code smells) and *Fowler and object-oriented both* facets. The PSs in *general* category are included in *unclear* facet. The distribution of PSs against these categories is listed in Figure 5c. Nearly half (49.3%) of the PSs worked on the refactoring activities proposed by Fowler [8] only, whereas only 7% of the PSs considered other object-oriented refactorings. There were 28 PSs (19.7%) which considered other object-oriented refactoring activities in conjunction with the refactoring activities proposed by Fowler [8].

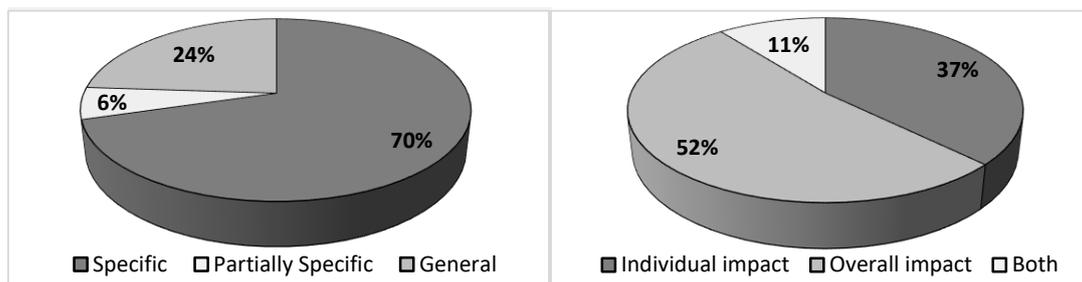

**Figure 9: Distribution of PSs according to - a) type of refactoring activities, and b) impact reported**

Furthermore, among the 108 PSs falling in *specific* and *partially specific* categories, a total of 154 distinct refactoring activities have been identified, out of which 70 are from the 72 refactoring activities proposed by Fowler [8]. The description of all 154 refactoring activities applied to measure the impact on software quality is provided online [36]. The space limitations preclude us from providing the detailed description of all these refactoring activities. Hence, the distribution of only 10 most used refactoring activities across selected PSs is shown in Figure 10. It can be observed that *Move Method, Extract Method* and *Extract Class* refactoring activities have been most frequently explored by the selected PSs. Each of the 81 (out of 154) refactoring activities were found to be investigated by more than one PS. The rest (47%) of the refactoring activities received lesser attention as these were considered by at most one PS each.

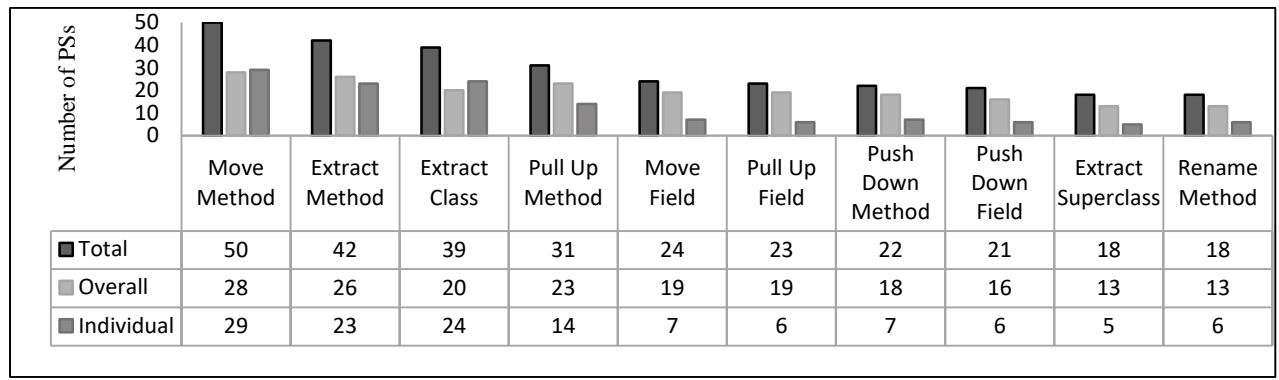

Figure 10: Distribution of 10 most applied refactoring activities

The selected PSs either reported the impact of individual refactoring activities or overall (known or unknown number of refactoring activities applied in combination on a software) refactoring activities, on the quality of software systems. Figure 9b shows that 74 PSs explored the impact of overall refactoring activities on software quality, whereas 53 PSs measured the impact of each individual refactoring activity on quality. There were 8 PSs ([S8], [S27], [S31], [S52], [S57], [S59], [S65] and [S73]) which partially addressed the effect of individual refactoring activities along with the impact of overall refactoring activities on software quality. The remaining 7 PSs have determined the impact of individual as well as overall refactoring activities on software quality. Furthermore, among 154 refactoring activities, 102 refactoring activities were classified based on their impact on software quality attributes. In addition, the individual effect of each of the rest 33.8% refactoring activities remained unexplored. The information of top ten refactoring activities concerning the total number of PSs that investigated a particular refactoring activity, the number of PSs that determined effect of the corresponding (*overall*) refactoring activity in conjunction with other refactoring activities, and the number of PSs that classified that (*individual*) refactoring activity based on its effect on software quality attributes is depicted in Figure 10.

*4.2.2. RQ2:* Which software quality measures have been used to study the impact of refactoring activities on software quality?

A software quality measure is the quantitative estimation of any feature or property of the software artifact. Based on Fenton and Pfleeger's [65] classification, we categorized the software quality measures into three subcategories namely product, process, and resource measures. The distribution of PSs among these categories is shown in Figure 11. A large proportion (62.7%) of the PSs considered only product measures to determine the impact of refactoring activities on software quality. Cohesion, coupling and complexity are among the most considered product quality measures. The others product quality measures listed under the *'others'* category include composition, polymorphism, encapsulation, messaging, etc. The process and resource measures were used by 17% and 6.3% of the PSs, respectively. Almost 6.3% of the PSs used both the product and process quality measures. Also, 3.5% of the PSs utilized product as well as resource measures and 2% of the PSs exploited process and resource measures. Only three PSs used all the product, process and resource measures to determine the impact of refactoring activities on software quality. This is understandable as it is not feasible to consider all product, process and resource measures together in one study as the choice of software measures depend upon the quality attributes and the research context targeted by that study.

The quality measures can be further categorized into *internal* and *external* measures. Figure 5d depicts the distribution of PSs among these categories. The selected PSs used internal measures to determine the impact of refactoring activities on internal quality attributes, or internal (as surrogate) and external measures to assess the impact of refactoring activities on external quality attributes. The internal quality measures were considered by 90 PSs. There were 16 PSs which accounted for both the internal and external quality measures. Among these 106 PSs, 71 studies used internal measures to assess the effect of refactoring activities on internal quality attributes. The rest 31 studies used internal measures as a means to determine the effect of refactoring activities on external quality attributes. In addition, four PSs considered internal quality measures to investigate the impact of refactoring on internal as well as external (as surrogates) quality attributes. The external quality measures were considered by approximately 36% of PSs.

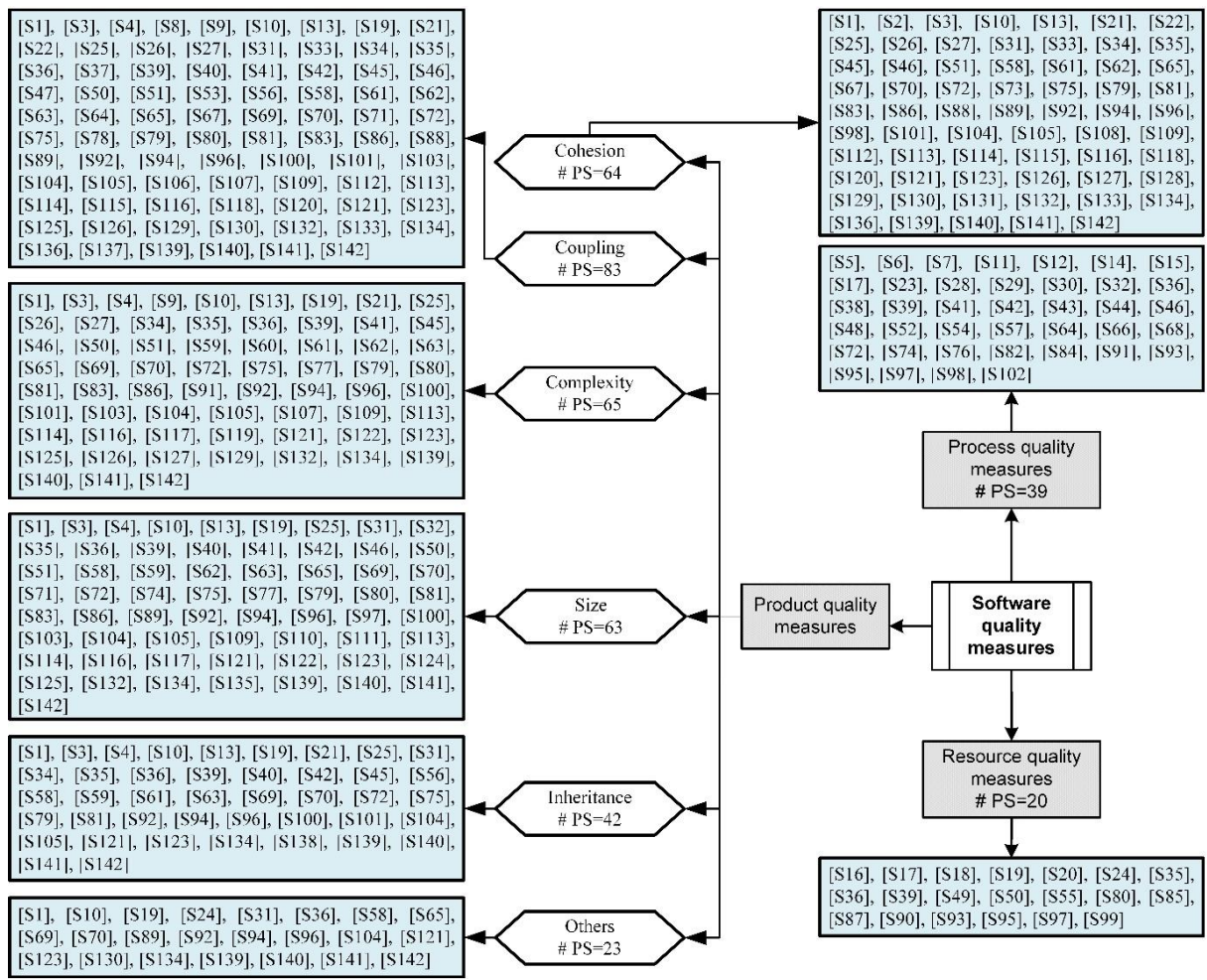

Figure 11: Classification of software quality measures

Figure 12 shows the number of PSs that used at least one measure from the well-known metric suites [57, 68]. The most commonly known metric suites include MOOSE [69-70], EMOOSE [71], L&K [72], Briand et al. [73], MOOD [74] and QMOOD [75] measures. The quality measures falling in multiple metric suites, like NOM metric, were not classified into any suite unless referenced by the researchers. Among 106 PSs that reported the use of internal quality measures, 74 PSs carried out their research on the measures from the aforementioned metric suites. Despite the other metric suites, MOOSE (also known as CK [68]) measures appeared in nearly half (37.3%) of the PSs. The *'others'* category includes the PSs that considered internal measures proposed by other researchers.

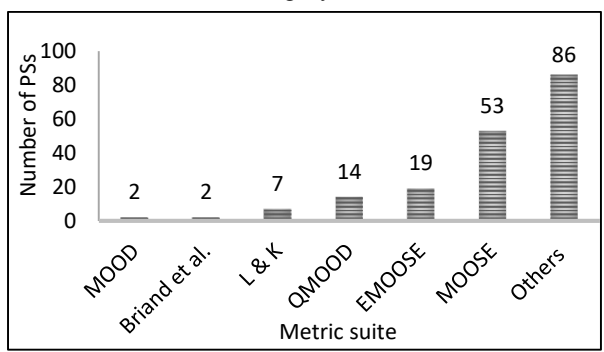

Figure 12: Metric suites considered in PSs

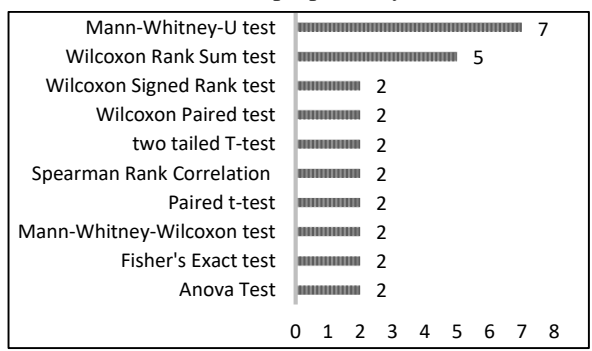

Figure 13: distribution of Statistical techniques exploited in selected PSs

A wide range (339) of quality measures has been employed by the selected PSs. The number of quality measures used by a PS varied from 1 to 32 [S39]. Table 9 lists the quality measures that are used in more than 15 PSs. The complete list including the remaining quality measures utilized by all 142 PSs is provided online [36]. Many of the PSs using LCOM did not mention the variant of LCOM metric (e.g. LCOM1, LCOM2, etc.). Therefore, in order to decide on the LCOM variant used in a given PS (that failed to report the type of considered LCOM measure in the full text of the article), we looked into the reference list of that PS. Table 9 shows that LOC, NOM, LCOM2 and CBO are among the most commonly used internal quality measures. Further, the major population (93.5%) of the quality measures were considered by fewer PSs (ten or less).

Table 9: Description of the most commonly used quality measures

| Acronym | Software Measure | Study References | #PSs |
|---|---|---|---|
| LOC | Lines of Code | [S3], [S10], [S13], [S19], [S25], [S32], [S35], [S36], [S39], [S40], [S41], [S46], [S50], [S51], [S59], [S62], [S63], [S65], [S72], [S74], [S75], [S77], [S79], [S80], [S81], [S83], [S86], [S92], [S97], [S100], [S109], [S110], [S111], [S113], [S114], [S116], [S124,] [S125] | 38 |
| NOM | Number Of Methods | [S3], [S4], [S21], [S13], [S25], [S26], [S39], [S41], [S42], [S50], [S51], [S59], [S62], [S65], [S70], [S75], [S80], [S81], [S83], [S92], [S94], [S96], [S101], [S104], [S107], [S109], [S113], [S114], [S116], [S121], [S123], [S129], [S132], [S134], [S139], [S140], [S141], [S142] | 38 |
| LCOM2 | Lack of Cohesion of Methods 2 | [S2], [S3], [S10], [S13], [S25], [S27], [S34], [S35], [S45], [S46], [S51], [S61], [S62], [S65], [S72], [S75], [S79], [S81], [S83], [S88], [S98], [S105], [S109], [S113], [S114], [S116], [S126], [S128], [S129], [S132], [S136] | 31 |
| CBO | Coupling Between Objects | [S3], [S9], [S10], [S13], [S21], [S25], [S34], [S35], [S40], [S42], [S45], [S46], [S51], [S56], [S61], [S65], [S69], [S72], [S75], [S78], [S79], [S80], [S81], [S86], [S105], [S107], [S114], [S125], [S136], [S137] | 30 |
| RFC | Response for a Class | [S3], [S9], [S10], [S13], [S21], [S25], [S26], [S34], [S35], [S42], [S45], [S46], [S51], [S61], [S65], [S67], [S72], [S75], [S81], [S83], [S86], [S88], [S101], [S103], [S105], [S107], [S109], [S129], [S136] | 29 |
| CC | Cyclomatic Complexity | [S13], [S19], [S21], [S27], [S36], [S39], [S50], [S51], [S59], [S60], [S61], [S62], [S63], [S65], [S69], [S79], [S80], [S83], [S86], [S91], [S100], [S117], [S119], [S122], [S125], [S126], [S132] | 27 |
| WMC | Weighted Methods per Class | [S3], [S9], [S10], [S25], [S26], [S27], [S34], [S35], [S45], [S46], [S61], [S65], [S71], [S72], [S75], [S79], [S81], [S83], [S86], [S92], [S101], [S103], [S105], [S126], [S127], [S129], [S132] | 27 |
| DIT | Depth of Inheritance Tree | [S3], [S10], [S13], [S19], [S25], [S34], [S35], [S36], [S39], [S40], [S42], [S45], [S56], [S59], [S61], [S72], [S75], [S78], [S79], [S81], [S101], [S105] | 22 |
| NOC | Number Of Children | [S3], [S10], [S13], [S21], [S25], [S34], [S35], [S42], [S45], [S56], [S61], [S72], [S75], [S79], [S81], [S92], [S101], [S105] | 18 |

Furthermore, it is noticed that software quality measures also play an eminent role while investigating the impact of refactoring activities on software quality. This is because after applying the same refactoring activity, a PS using one quality measure resulted in the abatement of a quality attribute, whereas another study using different quality measures caused an improvement in the same quality attribute (for example [S73]). The complete information about the association of software quality measures to quality attributes in the context of refactoring application is presented online [36].

*4.2.3. RQ3:* Which tools have been reported to predict or assess the impact of refactoring activities on software quality?

Only eleven PSs ([S34], [S56], [S70], [S88], [S92], [S94], [S104], [S118], [S121], [S130] and [S131]) reported tools to predict/assess the effect of refactoring activities on software quality. Refactoring impact prediction tools aid the developers in making design decisions and culling between different refactoring alternatives. The distribution of software quality attributes considered by the aforementioned tools is shown in Figure 14. Higo et al. [S34] implemented their proposed approach as a software tool to estimate the effect of nine refactoring activities on coupling, complexity, cohesion and inheritance quality attributes. The tool measures the CK metrics [69] for the original and refactored versions of the software systems, and compares these values to classify the impact of applied refactoring activities as positive or negative on internal quality attributes.

Sahraoui et al. [S56] proposed *OO1* tool to estimate the impact of three refactoring activities on software quality through a set of inheritance and coupling measures. Neto et al. [S70] provided an agent-based platform, named *AutoRefactoring* [76] to autonomously perform the refactoring of Java software systems. The proposed tool measures the flexibility, effectiveness, extensibility and reusability attributes of the original and refactored versions of the code. Later, it compares the calculated values of these quality attributes to determine the impact of refactoring activities on software quality. Jiau et al. [S88] implemented *OBEY* as an Eclipse plug-in, which executes batched refactoring plan on Java source code, and analyses the impact of refactoring on cohesion and coupling quality attributes. Mohan and Greer [S92] provided a search-based fully automated refactoring tool named as *MultiRefactor* which can perform 26 refactoring activities on Java source code. The proposed tool utilizes 23 software quality measures to determine the impact of refactoring on cohesion, coupling, complexity, composition, inheritance, encapsulation, polymorphism, messaging and size quality attributes.

O'Keeffe and O. Cinnéide [S94, S104] developed a prototype tool called *CODe-Imp* that takes Java source code as an input and outputs the refactored source code along with software quality improvement report including metric information and quality change. Tsantalis and Chatzigeorgiou [S130] provided an Eclipse plug-in named *JDeodorant* to automatically identify and apply *Move Method* refactoring opportunities with the aim to remove Feature Envy code smell. The proposed tool predicts the possible gains of applying all refactoring suggestions by measuring the value of entity placement metric employing cohesion and coupling measures. This tool was later extended to identify, apply and estimate the effect of *Extract Method* [S131] and *Extract Class* [S118] refactoring activities on cohesion and combination of quality attributes, respectively. Lee et al. [S121] implemented a tool called *Refactoring Scheduler* to identify an appropriate refactoring sequence that maximize software quality and remove software clones. The proposed tool estimates the impact of refactoring on combination of attributes before its actual application.

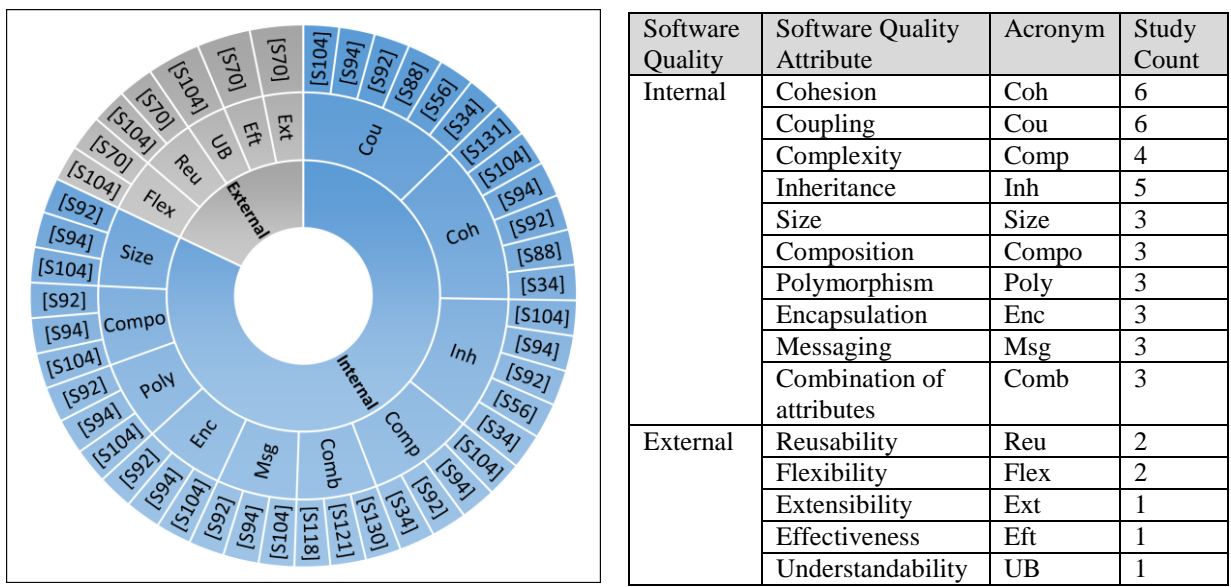

| Software Quality | Software Quality Attribute | Acronym | Study Count |
|---|---|---|---|
| Internal | Cohesion | Coh | 6 |
| | Coupling | Cou | 6 |
| | Complexity | Comp | 4 |
| | Inheritance | Inh | 5 |
| | Size | Size | 3 |
| | Composition | Compo | 3 |
| | Polymorphism | Poly | 3 |
| | Encapsulation | Enc | 3 |
| | Messaging | Msg | 3 |
| | Combination of attributes | Comb | 3 |
| External | Reusability | Reu | 2 |
| | Flexibility | Flex | 2 |
| | Extensibility | Ext | 1 |
| | Effectiveness | Eft | 1 |
| | Understandability | UB | 1 |

**Figure 14: Distribution of software quality attributes addressed by proposed tools**

*4.2.4. RQ4:* What datasets were used to conduct the empirical studies investigating the impact of refactoring activities on software quality?

A wide range of datasets (software systems) have been used to empirically evaluate the selected PSs. After going through 142 PSs, it has been observed that there were 26 PSs with insufficient details about the used software systems. Three PSs ([S4], [S14] and [S36]) did not report the exact number of datasets utilized to determine the effect of refactoring activities on software quality. Consequently, such datasets could not be included in the population of total datasets. The total number of software systems considered in the selected PSs were 472, out of which 294 were distinct. The distribution of software systems explored in selected PSs is shown in Figure 15a. A large portion of PSs (46.5%) used only one software system, and more than five subject systems have been very rarely used in selected PSs (24 out of 142). The maximum number of datasets considered in a PS were 23 [S79].

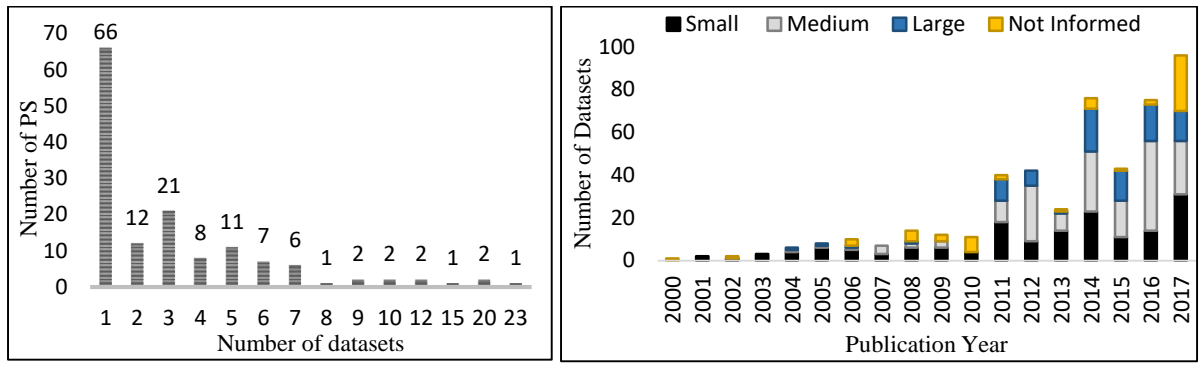

**Figure 15: Distribution of a) software systems used in the PSs b) datasets over publication year**

Among 294 distinct datasets, 15 datasets were utilized by more than three PS. The description of these datasets along with their respective PSs is enlisted in Table 10. JHotDraw has been used by a majority of PSs (12%). Figure 16 represents the set of PSs that used same datasets to determine the impact of refactoring activities on the same set of software quality attributes. The investigation of effect of refactoring activities on cohesion attribute utilizing GanttProject has received the (eleven PSs) higher researchers' attention. Following this, impact of refactoring on coupling employing GanttProject and JHotDraw, and cohesion using JHotDraw was determined by ten PSs each. The rest of the dataset-quality attributes pairs have been considered by less than ten PSs. Furthermore, among this distribution, most of the PSs ([S2, S3], [S8, S47], [S28, S38], [S38, S68], [S33, S120], [S26, S27, S83, S141, S142], [S8, S33, S106, S115, S130], etc.) targeting same datasets, used common refactoring activities.

*4.2.4.1. Dataset size*

The distribution of datasets with respect to their size over publication years (shown in Figure 15b) indicates that the number of datasets employed by the selected PSs increased corresponding to the increase in the number of PSs per publication year (presented in Figure 7). Since 2008, as the number of PSs were found less in number for the years 2013 and 2015 consequently, the number of datasets utilized by the PSs were less for these years. From 2011 onwards, it is noted that each of 65.3% of the selected PSs used more number of datasets (greater than one) to empirically investigate the impact of refactoring on software quality. Furthermore, an interesting observation is that recent PSs (2011 onwards) employed more number of medium and large datasets in comparison to earlier studies.

**Table 10: Description of data sets explored in PSs**

| Dataset name | Dataset Type | Programming Language | Studies | Count |
|---|---|---|---|---|
| JHotDraw | Open Source | Java | [S27], [S64], [S73], [S80], [S92], [S109], [S111], [S112], [S113], [S114], [S118], [S121], [S127], [S129], [S140], [S141], [S142] | 17 |
| GanttProject | Open Source | Java | [S26], [S27], [S67], [S73], [S74], [S83], [S89], [S109], [S112], [S113], [S114], [S116], [S139], [S140], [S141], [S142] | 16 |
| Apache Ant | Open Source | Java | [S7], [S8], [S15], [S29], [S52], [S78], [S79], [S83], [S93], [S98], [S101], [S119], [S138], [S139], [S141], [S142] | 16 |
| JEdit | Open Source | Java | [S8], [S15], [S33], [S38], [S47], [S68], [S70], [S88], [S106], [S112], [S115], [S120], [S130] | 13 |
| Xerces | Open Source | Java | [S7], [S70], [S74], [S79], [S98], [S113], [S119], [S138], [S139], [S140], [S141], [S142] | 12 |
| ArgoUML | Open Source | Java | [S7], [S11], [S13], [S54], [S68], [S79], [S98], [S113], [S114], [S139] | 10 |
| JFreeChart | Open Source | Java | [S74], [S115], [S121], [S130], [S131], [S138], [S140], [S141], [S142] | 9 |
| Eclipse | Open Source | Java | [S11], [S23], [S38], [S58], [S88], [S101], [S113], [S114], [S133] | 9 |
| JUnit | Open Source | Java | [S68], [S71], [S79], [S86], [S103] | 5 |
| Columba | Open Source | Java | [S27], [S28], [S33], [S38], [S120] | 5 |
| Apache JMeter | Open Source | Java | [S52], [S106], [S119], [S137], [S138] | 5 |
| Antlr | Open Source | Java | [S27], [S71], [S86], [S103], [S121] | 5 |
| Xalan | Open Source | Java | [S8], [S15], [S119], [S138] | 4 |
| JabRef | Open Source | Java | [S73], [S95], [S97], [S109] | 4 |

| ArtOfIllusion | Open Source | Java | [S73], [S109], [S140], [S141] | 4 |

Moreover, it is analysed that 33.9%, 35.6% and 18.4% of the total 472 systems were small, medium and large, respectively. The sizes of 57 (12.1%) datasets were not described clearly. Furthermore, the largest datasets in the unbounded *large* category were Eclipse [S114], Mozilla [S11] and a commercial project [S30] with 23462 classes (1710 KLOC), 12358 files (3169 KLOC) and 7489 classes (266,629 LOC), respectively. Figure 17 depicts the distribution of the PSs across aforementioned categories of dataset sizes, viz. small, medium, and large. It is analysed from the data in Figure 17 that 38.7%, 13.4% and 4.9% of the selected PSs used small, medium and large datasets, respectively. Variable sized software systems were utilized in 30.3% of the PSs. Almost 2.8% and 12.7% of the PSs partially and completely do not specify the dataset sizes, respectively.

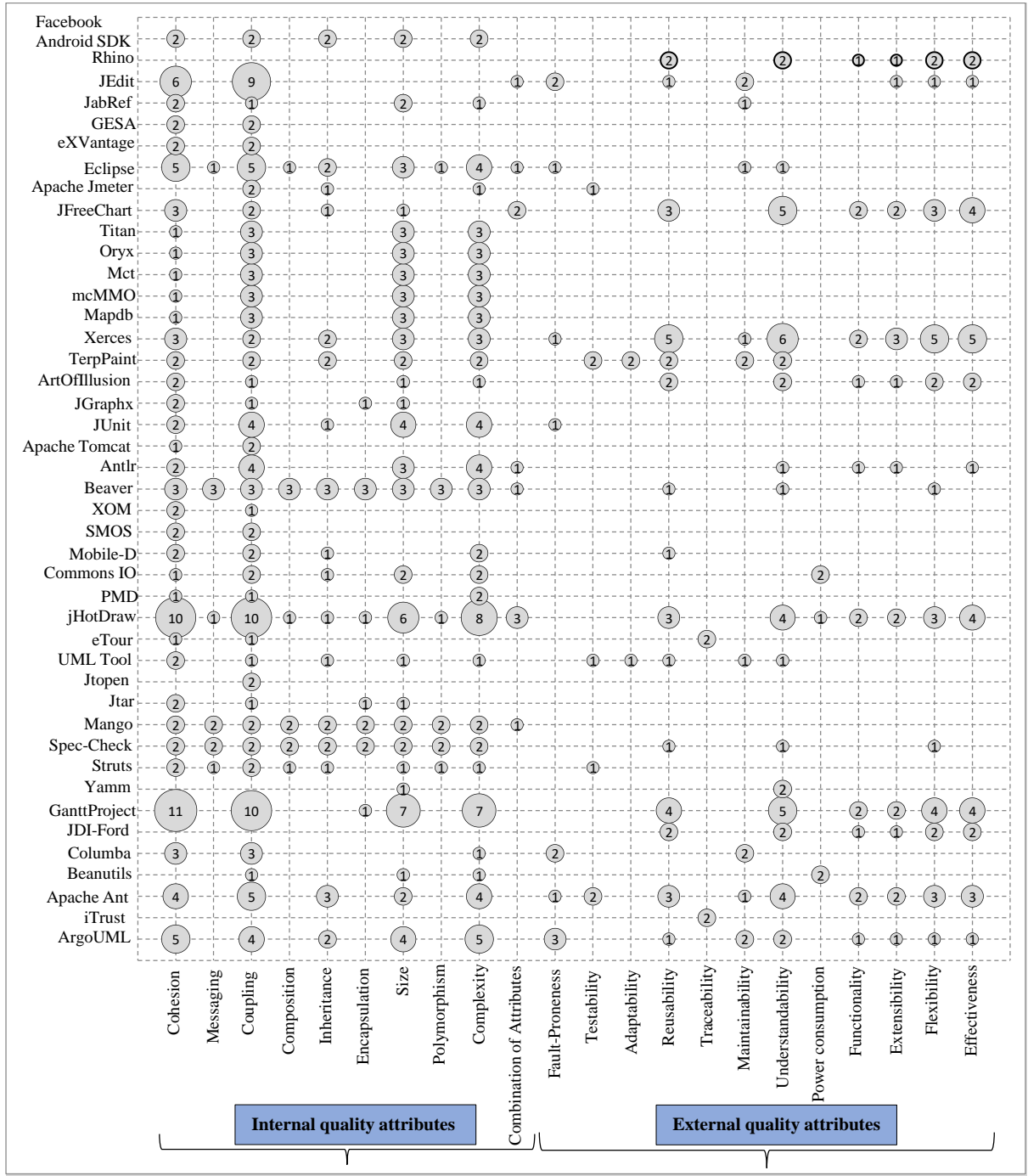

**Figure 16: Distribution of PSs using datasets by quality attribute**

*4.2.4.2. Dataset programming language*

Among 472 datasets, the majority of datasets (85.6%) were written in Java programming language. The datasets implemented in C++ (5.1%), C# (4.2%), Python (0.2%) and Swift (0.2%) were used infrequently by the researchers. The programming languages of 4.7% of the datasets were not stated. The distribution of PSs across these programming languages is depicted in Figure 17. Out of 142 PSs, 0.7% and 4.9% of the PSs partially and completely failed to report the programming language of the datasets, respectively.

*4.2.4.3. Dataset type*

A large portion of datasets (76.1%) are open source systems. The rest of the datasets are academic (7.8%), commercial (10.4%) and student (3.8%) projects. In addition, the dataset type was not found for 1.9% of datasets. With respect to PSs, open source, academic, commercial and student projects were used in 55.6%, 14.8%, 14.8% and 3.5% of the PSs, respectively. Further, 8.4% of the PSs used more than one type of datasets. Figure 17 shows the distribution of PSs among these dataset types.

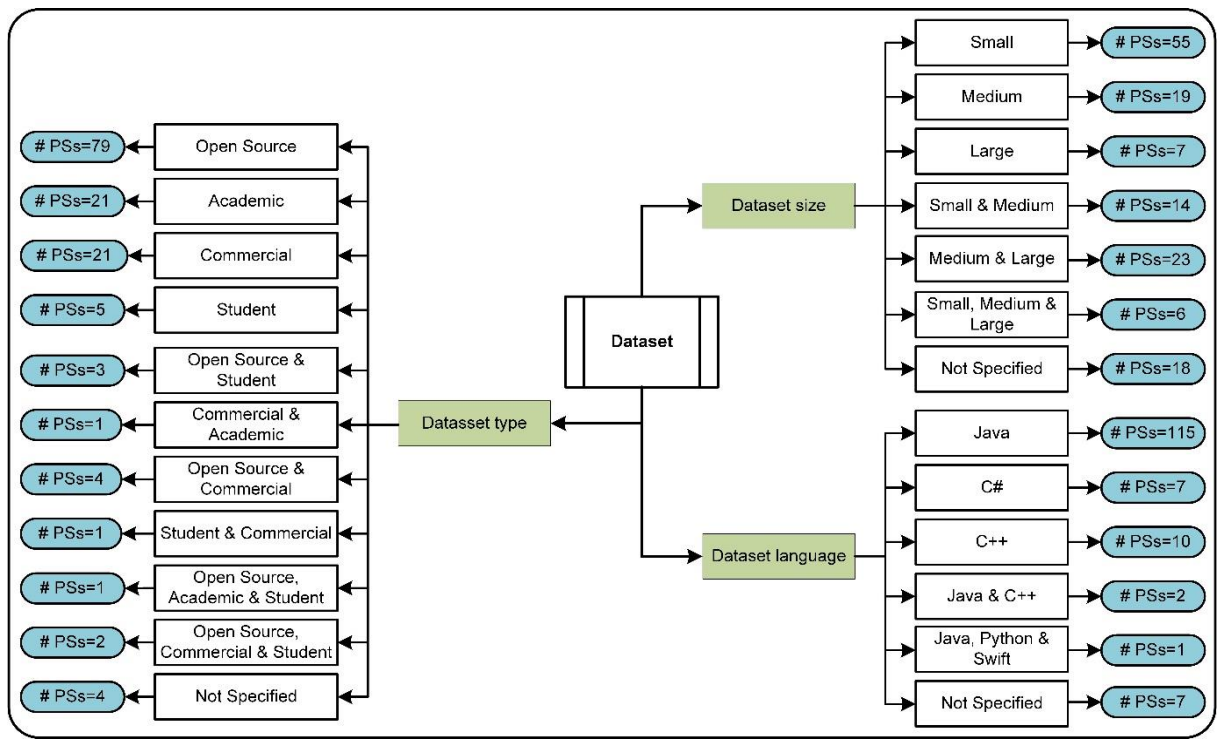

**Figure 17: Distribution of PSs according to size, programming language and dataset type**

*4.2.5. RQ5:* What is the current state of knowledge about the impact of refactoring activities on software quality?

A large number of refactoring activities (154) and quality attributes (37) have been considered in selected PSs. The distribution of PSs involving the use of most frequently explored refactoring activities and quality attributes is depicted in Figure 18. The impact of *Move Method* and *Extract Class* refactoring activities on coupling attribute was addressed by approx. 14% and 12% of the PSs, respectively, which is higher than that for any other refactoring activity-quality attribute pair. Following this, effect of *Move Method* on cohesion, and *Extract Class* refactoring activities on cohesion as well as complexity attribute has been studied in 15 PSs each. The rest of refactoring activity-quality attribute pairs have been considered by less than 15 PSs each.

Out of 37 software quality attributes considered by selected PSs, ten are internal and 27 are external attributes. Figure 19 depicts that cohesion, coupling, complexity, size and inheritance are the most explored quality attributes with PS counts of 53, 69, 54, 50 and 28, respectively. Polymorphism and messaging attributes have received very little attention (five PSs each). Furthermore, we can observe that understandability, maintainability and reusability are amongst the three most frequently investigated external quality attributes. About 52% of the external quality attributes have been covered by fewer PSs (four or less), indicating that researchers generally prefer to target and investigate just a few external attributes in one study.

We applied vote-counting approach while combining the empirical results related to the effect of refactoring on software quality because only relatively few (26 PSs) PSs applied statistical techniques to analyse the statistical significance of obtained outcomes. Among these 26 PSs, 17 distinct statistical approaches were considered. The distribution of statistical methods utilized by more than one PS is depicted in Figure 13. It can be observed from Figure 13 that Mann-Whitney-U test is exploited by maximum number of PSs. Following this, five PSs used Wilcoxon Rank Sum test. Each of the other statistical techniques were considered by less than two PSs. In addition, remaining 81.7% PSs simply reported results based on the difference in the values of software quality measures obtained before and after the application of software refactoring activities, rather than supporting their results through the statistical significance of change in software quality measure values.

Furthermore, these 26 PSs explored the significance of reported outcomes by employing the statistical techniques at two significance levels of p-values namely 0.05 and 0.01. However, among these 26 PSs, only eight PSs reported the study outcomes at p-value of 0.01. Therefore, similar to Dallal and Abdin [21], we too combined the findings reported at 0.01 and 0.05 significance levels of p-values. Furthermore, we adopted the categorization used by Dallal and Abdin [21] to divide the study results into six categories namely *positive, significant positive, negative, significant negative, unchanged* and *insignificantly changed*. The results reported by only 26 PSs, which applied statistical methods, are placed in *significant positive, significant negative* and *insignificantly changed* categories. Tables 11 and 12 represent the vote-counting results concerning the impact of overall refactoring activities on internal quality and external quality, respectively. The vote-counting results for the effect of individual refactoring activities on internal and external quality attributes are provided online [36]. The vote-counting results under +IQ/+EQ, ++IQ/++EQ, -IQ/-EQ, --IQ/--EQ, =IQ/=EQ and !IQ/!EQ columns represent *positive, significant positive, negative, significant negative, unchanged* and *insignificantly changed* votes, respectively. In addition, the corresponding PSs against *positive, significant positive, negative, significant negative, unchanged* and *insignificantly changed* categories are enlisted underneath +PSs, ++PSs, -PSs, --PSs, =PSs and !PSs columns, respectively.

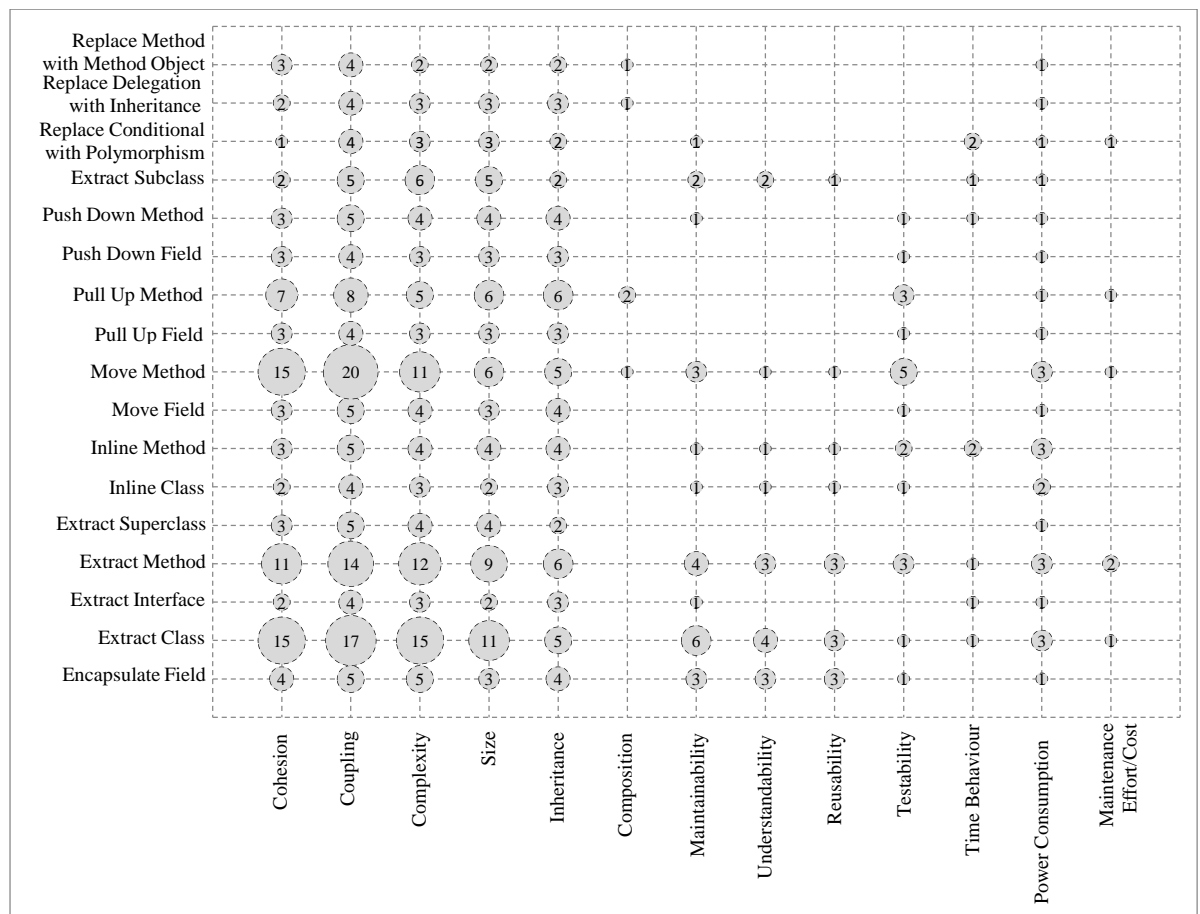

**Figure 18: Distribution of PSs depicting the impact of frequently considered refactoring activities on most explored quality attributes**

Based on the approach followed by Dallal and Abdin [21], the number of votes of a PS for a particular attribute are calculated by multiplying the number of quality measures, refactoring activities and datasets employed by that PS. For example, a study [S25] determined the impact of 12 refactoring activities on cohesion using nine quality measures and six datasets. The total number of votes of this study [S25] for cohesion will be equal to $12 \times 9 \times 6 = 648$. Furthermore, the number of refactoring activities for the studies investigating the effect of overall refactoring activities on software quality are considered as one during vote-counting. This is because these studies report single combined impact outcome of all known or unknown number of refactoring activities. Moreover, similar to Dallal and Abdin [21], we also believe that the significant impact findings reported by the studies employing statistical techniques are more reliable than other non-significant findings. Therefore, if the statistically significant votes exist for a quality attribute, the final conclusions were made solely on the basis of these significant results and the other impact results were ignored. The non-significant votes were considered only for those quality attributes where not even a single PS explored the significance of reported outcomes.

Figures 20, 21, and 22 summarize the findings concerning the final effect of overall refactoring activities on internal quality, external quality, and individual refactoring activities on software quality, respectively. A link to +E represents *positive* or *significant positive* effects of refactoring activities on particular internal or external quality attributes. The connection to -E indicates *negative* or *significant negative* effect of refactoring activities on respective internal or external quality attributes. An association to =E signifies that the relevant studies did not identify any definite impact of refactoring activities on corresponding internal or external quality attributes. A link to ~E represents an inconsistent effect of refactoring activities on internal or external software quality attributes. Furthermore, the third, fourth and fifth columns of Figures 20 and 21 represent the total number of PSs, distinct datasets and percentage of significant or non-significant votes corresponding to each quality attribute, respectively. The number of studies conducted in academic and industrial settings are shown in different colours.

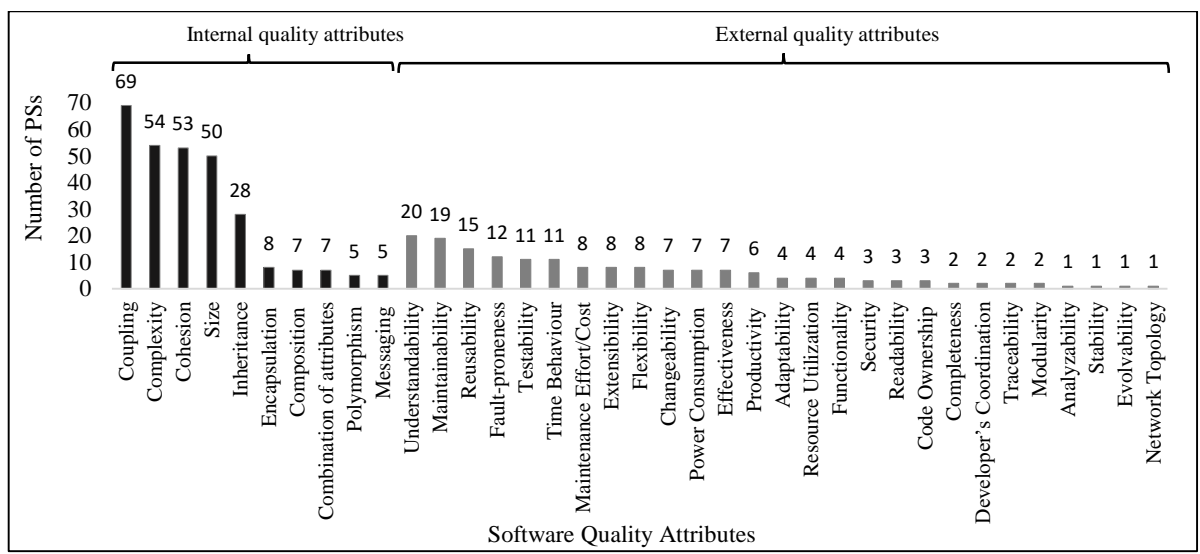

**Figure 19: Number of PSs addressing software quality attributes**

The *significant positive* (+E) impact of refactoring activities on a particular internal/external quality attribute in a particular setting (academic/industrial) is calculated by dividing the 'number of votes corresponding to the *significant positive* impact' by 'the total number of significant votes (sum of *significant positive, significant negative* and *insignificantly changed*)' for that quality attribute. We used similar approach for other cases, viz. *positive, negative, significant negative, unchanged* and *insignificantly changed*. The final impact of a refactoring activity on a software quality attribute is decided based on 50% threshold value [21]. If the threshold value on all sides i.e. positive, neutral or negative is below 50%, it becomes difficult to indicate whether the overall quality would be increased or decreased at the end. Hence, we call these cases as inconsistent (~E). For example, if out of 40 significant votes for studies conducted in an academic setting, 8, 14 and 18 votes represent *significant positive, insignificantly changed,* and *significant negative* effect of refactoring on cohesion, respectively. Then the percentage of positive, neutral and negative impact comes out to 20%, 35% and 45%, respectively. The final impact of refactoring on cohesion will be inconsistent as threshold value in all cases is below 50%.

*4.2.5.1. Impact of overall refactoring activities on internal quality attributes*

A total of ten internal quality attributes have been investigated in 142 PSs. These quality attributes along with their respective final impact are enlisted in Figure 20. Most of the studies used coupling, cohesion, complexity, inheritance and size attributes to quantify the effect of refactoring activities on internal quality attributes. The other internal quality attributes found in this SMS are considered by fewer PSs (eight or less).

Furthermore, the researchers validated their studies either in industrial or academic settings. There is huge difference in findings between the studies employing industrial and academic settings. The vote-counting results for each quality attribute are enlisted in Table 11, and the final impact of refactoring activities on each internal quality attribute calculated from these votes is depicted in Figure 20. Among 53 PSs focusing on determining the impact of overall refactoring activities on internal quality, a majority of the studies performed in academic settings found that the application of refactoring activities on object-oriented software affected the internal quality attributes in a positive way. Among ten internal quality attributes, the academic studies reported the positive effect of refactoring activities on seven quality attributes namely coupling, size, encapsulation, composition, messaging, polymorphism and combination of attributes with 63%, 75%, 78%, 64%, 82%, 55% and 89% positive or significant positive votes, respectively. In addition, the impact of refactoring activities on cohesion, complexity and inheritance attributes is found to be inconsistent. It is to be noted here that for cohesion, coupling, complexity and size, only the significant votes were considered while calculating the final impact. On the other hand, non-significant votes were taken into account for remaining internal quality attributes because of the non-existence of significant votes.

A diverse trend is observed for the studies conducted in industrial settings as studies with almost 100%, 67%, 100%, 67% and 100% unchanged or insignificantly changed votes did not identify any definite impact of refactoring activities on cohesion, size, inheritance, encapsulation, and composition attributes, respectively. For industrial studies, we considered significant votes for cohesion, coupling, complexity, size and inheritance attributes as these significant votes exist for these quality attributes only. Furthermore, the industrial authors found the inconsistent impact of refactoring on coupling and complexity attributes. This provides an interesting observation *that studies conducted in academic settings found more positive impact of refactoring on software quality than studies performed in industries*. However, this should be viewed in the light of the fact that the number of industrial studies (17 PSs) are less as compared to academic studies (125 PSs). Furthermore, the effect of refactoring activities on messaging, polymorphism and combination of attributes has not been evaluated in industrial settings, whereas academic researchers reported the positive impact of refactoring on aforementioned quality attributes.

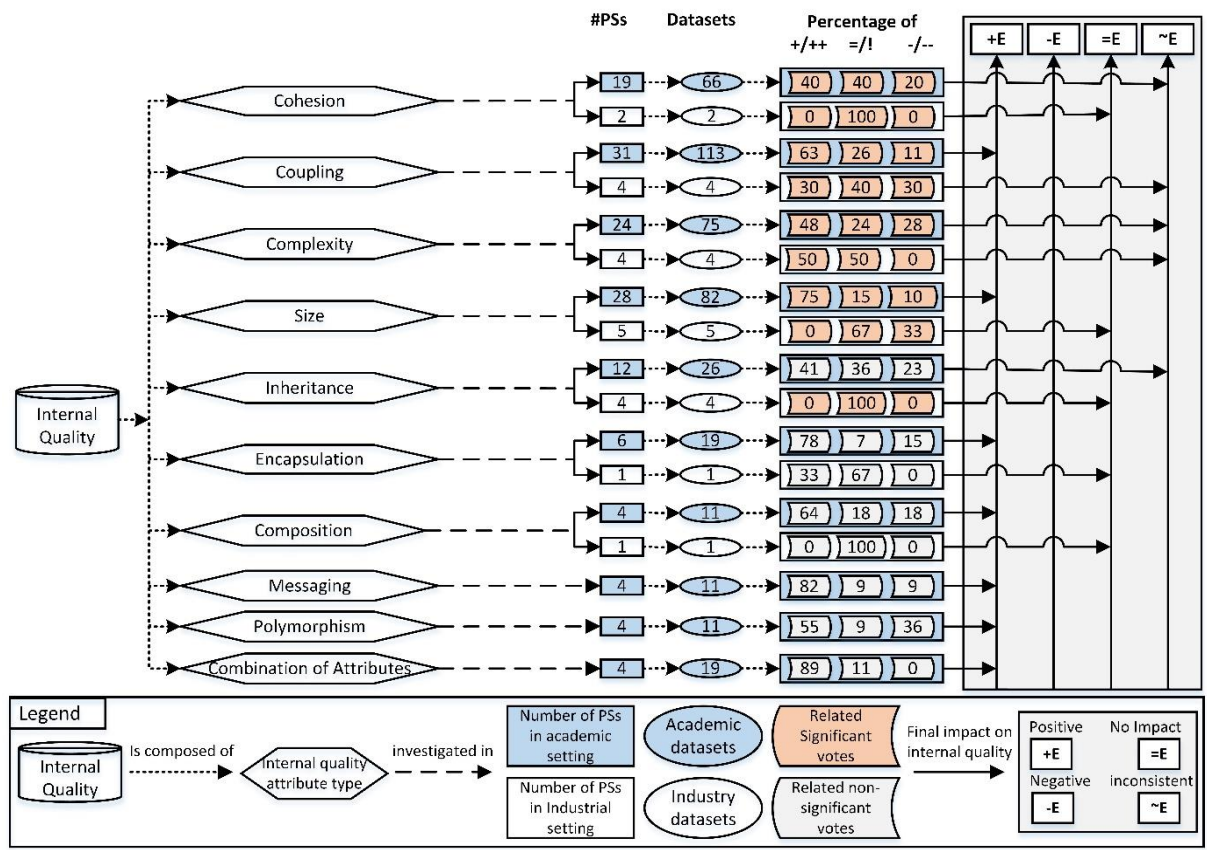

**Figure 20: The impact of refactoring activities on internal quality attributes**

In order to compare our findings with Dallal and Abdin [21], we have also reported the combined overall results irrespective of whether the study is performed in academic or industrial settings. The combined vote-counting results of studies investigating the effect of refactoring on internal quality are provided online [36]. In general, the application of refactoring resulted in the improvement of coupling, size, encapsulation, messaging, polymorphism, composition and combination of attributes with 58%, 69%, 73%, 82%, 55%, 58% and 89% positive or significant positive votes, respectively. Furthermore, cohesion and inheritance attributes remains unchanged and complexity attribute is affected in an inconsistent way, following the application of refactoring activities. Several different outcomes were reported by Dallal and Abdin [21] mentioning inconsistent impact on cohesion, size, inheritance, encapsulation and composition, positive effect on complexity and negative impact of refactoring on polymorphism attribute. Moreover, we found similar outcomes with Dallal and Abdin [21] only in case of coupling attribute.

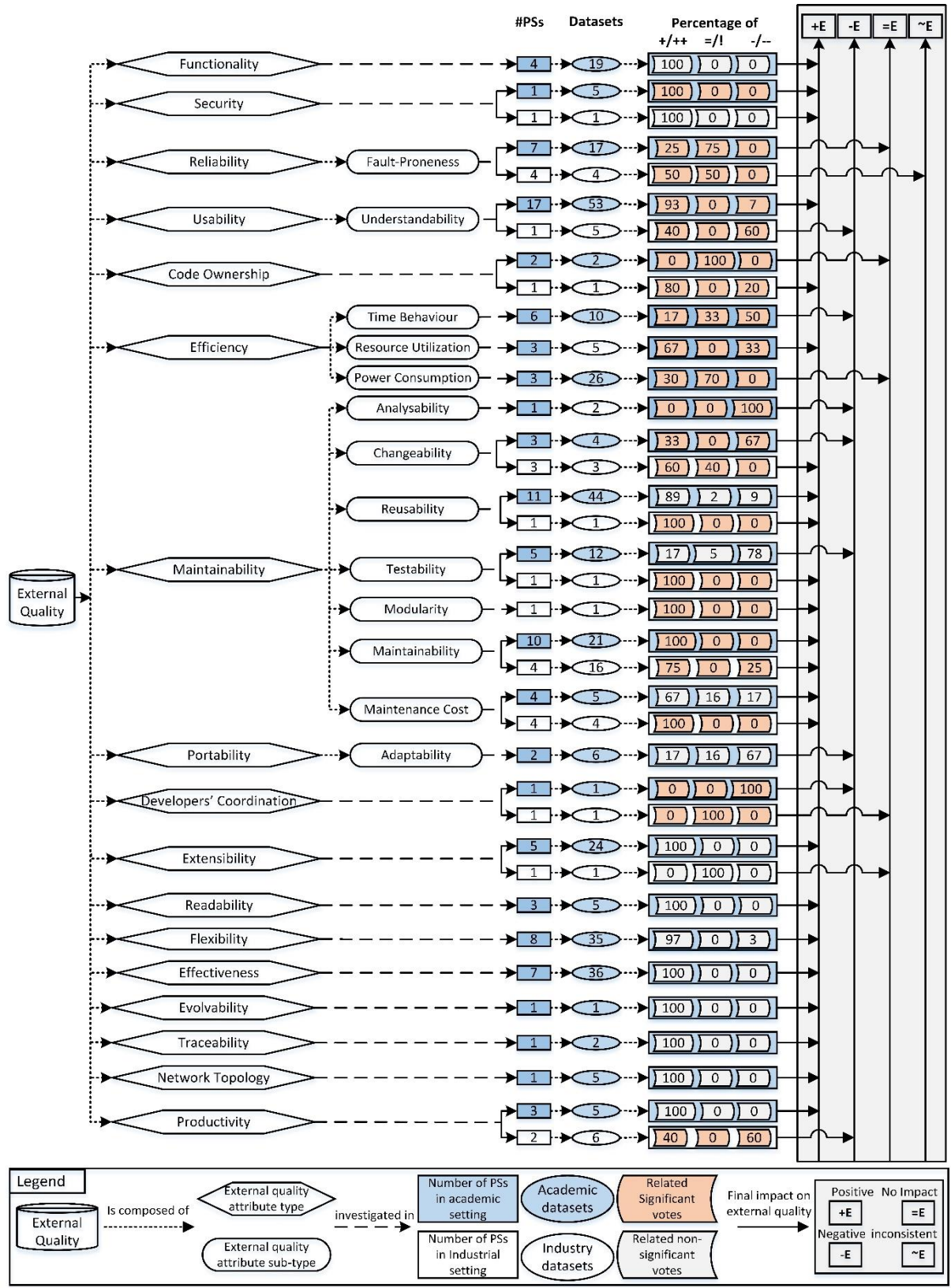

**Figure 21: The impact of overall refactoring activities on external quality attributes**

*4.2.5.2. Impact of overall refactoring activities on external quality attributes*

A wide range of external quality attributes have been considered across 142 PSs, as shown in Figure 19. The investigation on 41% of the external quality attributes is quite sparse (three or less PSs per attribute). The results for the impact of overall refactoring activities on external quality attributes is presented in Figure 21. The international standard ISO/IEC 25010 [77] was used to map the external quality attribute subtypes to most appropriate external quality attributes found in PSs. The final impact of overall refactoring activities on external quality attributes is calculated from the vote-counting results represented in Table 12. It is to be noted here that we considered non-significant vote-counting results for only those quality attributes (e.g. extensibility, readability, flexibility, etc.) whose significant vote results were missing. Furthermore, it can be observed from Figure 21 that the outcomes obtained from the studies conducted in academic settings are not identical to those performed in industrial settings except for security, reusability, maintenance cost/effort and maintenance quality attributes.

For PSs conducted in academic settings, the application of refactoring activities results in improvement of functionality, security, understandability, resource utilization, reusability, maintainability, maintenance cost/effort, extensibility, readability, flexibility, effectiveness, evolvability, traceability, network topology, and productivity attributes with 100%, 100%, 93%, 67%, 89%, 100%, 67%, 100%, 100%, 97%, 100%, 100%, 100%, 100% and 100% votes, respectively. Also, the percentages of votes showing neutral impact of refactoring activities on fault-proneness, code ownership and power consumption are 75%, 100% and 70%, respectively. The effect of refactoring on time behaviour, analysability, changeability, developers' coordination, testability and adaptability is negative with 50%, 100%, 67%, 100%, 78% and 67 % votes, respectively.

Furthermore, academic authors did not evaluate modularity attribute in academic setting. However, the observed trend for the studies performed in industrial settings is somewhat different. Industry authors reported positive impact of refactoring activities on security, code ownership, changeability, reusability, testability, modularity, maintainability and maintenance effort/cost; and negative effect on understandability and productivity attributes. In addition, we found neutral impact across the industrial studies for developers' coordination and extensibility quality attributes; and inconsistent effect of refactoring on fault-proneness attribute. Almost 12 external quality attributes are not evaluated in industrial setting.

Moreover, we also reported the combined impact of refactoring on software quality irrespective of setting in which the study is conducted in order to compare our findings with Dallal and Abdin [21]. In general, refactoring activities caused all the external quality attributes to improve or degrade except for fault-proneness and power consumption attribute, which resulted in insignificant change (=EQ) in quality. The vote-counting results concerning the combined effect of overall refactoring on software quality irrespective of study setting are provided online [36]. Furthermore, the refactoring activities have positive or significant positive impact on changeability, effectiveness, evolvability, extensibility, flexibility, network topology, readability, traceability, code ownership, maintainability, maintenance effort/cost, modularity, resource utilization, reusability, security, testability, and understandability attributes; and negative or significant negative impact on adaptability, time behaviour, developers' coordination, productivity and analysability attributes. The reported findings of this SMS are similar to Dallal and Abdin [21] only for flexibility, extensibility, effectiveness, adaptability, functionality, maintainability, modularity, and fault-proneness quality attributes.

*4.2.5.3. Impact of individual refactoring activities on software quality*

About 42% of the PSs reported the impact of individual refactorings on software quality attributes. These PSs provided the classification of 102 (66.2%) refactoring activities according to their effect on software quality attributes. This classification resulted in 661 refactoring activity-quality attribute pairs. The space limitations preclude us from providing the detailed description of each refactoring activity-quality attribute pair. Also, among these 661 pairs, 590 refactoring activity-quality attribute pairs were investigated by few PSs i.e. each of the 500 and 90 pairs by one and two PSs, respectively. These scarcely investigated pairs may not provide a strong evidence to support the effect of a particular refactoring activity on software quality. Hence, only the most frequently considered 71 refactoring activity-quality attribute pairs are shown in Figure 22. However, the complete information regarding the aforementioned classification is provided online [36]. We did not differentiate between the outcomes of PSs conducted in academic and industrial settings, because only three ([S37], [S46] and [S51]) studies reported the effect of each individual refactoring activity on software quality in industrial settings.

For the refactoring activity-quality attribute pairs depicted in Figure 22, it can be observed that most of the refactoring activities have either negative or neutral impact on software quality attributes. The application of each refactoring activity resulted in an abatement of testability, maintainability and understandability attributes. The *Reverse Conditionals* refactoring activity has no impact on reusability attribute. Furthermore, the influence of each refactoring activity on inheritance in also found as neutral except for *Extract Class* and *Extract Interface* refactoring activities. Furthermore, it can be noted from Figure 22 that if an individual refactoring activity resulted in the improvement of some quality attributes, it caused the abatement of several other quality attributes simultaneously. Hence, no particular trend was found regarding the impact of each refactoring activity on internal and external quality attributes. The detailed impact of each refactoring activity on every quality attribute is provided online [36] due to space limitations.

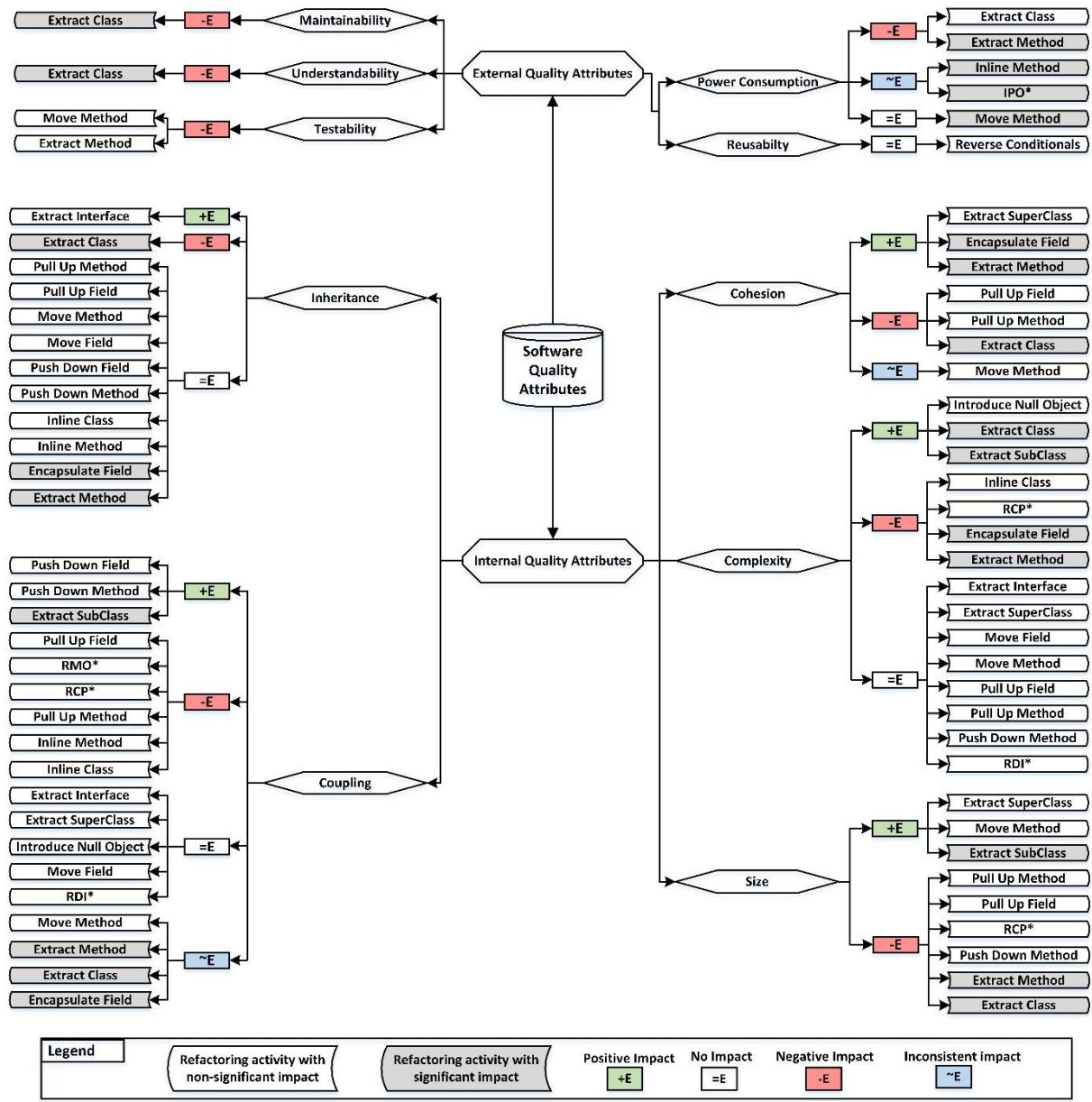

\* IPO: Introduce Parameter Object; RCP: Replace Conditional with Polymorphism; RMO: Replace Method with Object; RDI: Replace Delegation with Inheritance

**Figure 22: The impact of individual refactoring activities on software quality**

Table 11: Vote counting results regarding the impact of overall refactoring on internal software quality attributes

| Quality Attribute | A/I* | ++PSs | ++IQ | +PSs | +IQ | !PSs | !IQ | =PSs | =IQ | -PSs | -IQ | --PSs | --IQ |
|---|---|---|---|---|---|---|---|---|---|---|---|---|---|
| Cohesion | A | [S86] | 2 | [S2], [S3], [S27], [S61], [S65], [S73], [S83], [S88], [S89], [S92], [S96], [S98], [S104], [S105] | 77 | [S86] | 2 | [S3], [S27], [S73], [S75], [S89], [S94] | 20 | [S3], [S27], [S35], [S61], [S67], [S73], [S75], [S94] | 39 | [S86] | 1 |
| | I | × | 0 | [S31] | 1 | [S45] | 1 | [S31] | 3 | × | 0 | × | 0 |
| Coupling | A | [S71], [S86], [S103] | 34 | [S3], [S9], [S19], [S27], [S35], [S41], [S53], [S56], [S61], [S64], [S65], [S67], [S75], [S78], [S83], [S88], [S89], [S92], [S104], [S105] | 82 | [S40], [S86], [S103], [S125] | 14 | [S8], [S27], [S47], [S75], [S78], [S83], [S89], [S94], [S105] | 35 | [S3], [S9], [S27], [S35], [S36], [S61], [S75], [S78], [S80], [S86], [S94], [S96] | 59 | [S40], [S71], [S86], [S125] | 6 |
| | I | [S39], [S45] | 3 | [S31] | 1 | [S39] | 4 | [S42] | 2 | [S42] | 1 | [S39] | 3 |
| Complexity | A | [S71], [S86], [S103], [S125] | 24 | [S3], [S19], [S27], [S35], [S41], [S59], [S60], [S61], [S65], [S75], [S80], [S83], [S92], [S94], [S104], [S105], [S117] | 114 | [S86], [S125] | 12 | [S27], [S61], [S75], [S80], [S83], [S92], [S94], [S96] | 44 | [S3], [S9], [S27], [S36], [S41], [S59], [S61], [S65], [S75], [S80], [S83], [S86], [S96] | 46 | [S71], [S86], [S103] | 14 |
| | I | [S39], [S45] | 2 | [S77], [S91] | 5 | [S45] | 2 | × | 0 | × | 0 | × | 0 |
| Inheritance | A | × | 0 | [S3], [S56], [S61], [S75], [S92], [S94], [S96], [S105] | 22 | × | 0 | [S3], [S19], [S36], [S75], [S92], [S94], [S96], [S104], [S105] | 19 | [S3], [S35], [S61], [S75], [S96], [S104] | 12 | × | 0 |
| | I | × | 0 | × | 0 | [S39], [S45] | 3 | [S31] | 4 | [S42] | 5 | × | 0 |
| Size | A | [S40], [S71], [S86], [S103] | 52 | [S3], [S10], [S35], [S41], [S65], [S74], [S75], [S83], [S89], [S92], [S94], [S96], [S97], [S104], [S105], [S110], [S111], [S117], [S124], [S135] | 45 | [S71], [S86], [S103] | 10 | [S75], [S89], [S92], [S94], [S96], [S110] | 14 | [S3], [S10], [S19], [S36], [S59], [S65], [S80], [S94], [S96], [S104], [S110], [S124] | 21 | [S71], [S86], [S103] | 7 |
| | I | × | 0 | [S31], [S42], [S77] | 4 | [S39] | 4 | × | 0 | [S32], [S42], [S77] | 9 | [S39] | 2 |
| Encapsulation | A | × | 0 | [S65], [S89], [S92], [S96] | 21 | × | 0 | [S89], [S94] | 2 | [S89], [S94], [S104] | 4 | × | 0 |
| | I | × | 0 | [S31] | 1 | × | 0 | [S31] | 2 | × | 0 | × | 0 |
| Composition | A | × | 0 | [S92], [S94], [S96] | 7 | × | 0 | [S94], [S96] | 2 | [S94], [S104] | 2 | × | 0 |
| | I | × | 0 | × | 0 | × | 0 | [S31] | 1 | × | 0 | × | 0 |
| Messaging | A | × | 0 | [S92], [S94], [S96], [S104] | 9 | × | 0 | [S94] | 1 | [S96] | 1 | × | 0 |
| Polymorphism | A | × | 0 | [S92], [S94] | 6 | × | 0 | [S94] | 1 | [S94], [S96], [S104] | 4 | × | 0 |
| Combination of Attributes | A | × | 0 | [S10], [S92], [S121], [S134] | 17 | × | 0 | [S10] | 2 | × | 0 | × | 0 |

Table 12: Vote counting results regarding the impact of overall refactoring on external software quality attributes

| Quality Attribute | A/I | ++PSs | ++EQ | +PSs | +EQ | !PSs | !EQ | =PSs | =EQ | -PSs | -EQ | --PSs | --EQ |
|---|---|---|---|---|---|---|---|---|---|---|---|---|---|
| **Adaptability** | A | × | 0 | [S3] | 1 | × | 0 | [S3] | 1 | [S3], [S75] | 4 | × | 0 |
| **Analysability** | A | × | 0 | × | 0 | × | 0 | × | 0 | × | 0 | [S36] | 2 |
| **Changeability** | A | [S30] | 1 | [S64] | 1 | × | 0 | × | 0 | × | 0 | [S36] | 2 |
| | I | [S39], [S84] | 3 | [S32] | 3 | [S84] | 2 | × | 0 | × | 0 | × | 0 |
| **Code Ownership** | A | × | 0 | [S93] | 2 | [S97] | 1 | × | 0 | × | 0 | × | 0 |
| | I | [S39] | 4 | × | 0 | × | 0 | × | 0 | × | 0 | [S39] | 1 |
| **Developers Coordination** | A | × | 0 | × | 0 | × | 0 | × | 0 | × | 0 | [S17] | 3 |
| | I | × | 0 | × | 0 | [S39] | 2 | × | 0 | × | 0 | × | 0 |
| **Effectiveness** | A | × | 0 | [S70], [S121], [S123], [S139], [S140], [S141], [S142] | 36 | × | 0 | × | 0 | × | 0 | × | 0 |
| **Evolvability** | A | × | 0 | [S53] | 1 | × | 0 | × | 0 | × | 0 | × | 0 |
| **Extensibility** | A | × | 0 | [S70], [S121], [S123], [S139], [S141] | 24 | × | 0 | × | 0 | × | 0 | × | 0 |
| | I | × | 0 | × | 0 | × | 0 | [S31] | 2 | × | 0 | × | 0 |
| **Fault-Proneness** | A | [S30] | 1 | [S28], [S54] | 7 | [S7] | 3 | × | 0 | [S14], [[S38], S68] | 7 | × | 0 |
| | I | [S39] | 1 | [S102] | 1 | [S84] | 1 | × | 0 | [S42] | 1 | × | 0 |
| **Flexibility** | A | × | 0 | [S70], [S96], [S104], [S123], [S139], [S140], [S141], [S142] | 34 | × | 0 | × | 0 | [S70] | 1 | × | 0 |
| **Functionality** | A | × | 0 | [S121], [S123], [S139], [S141] | 19 | × | 0 | × | 0 | × | 0 | × | 0 |
| **Maintainability** | A | [S11], [S97] | 4 | [S3], [S19], [S35], [S36], [S59], [S65], [S98] | 12 | × | 0 | [S3] | 1 | [S3], [S75] | 4 | × | 0 |
| | I | [S39] | 3 | [S63], [S69], [S100] | 13 | × | 0 | × | 0 | [S69], [S100] | 2 | [S39] | 1 |
| **Maintenance Cost/ Effort** | A | × | 0 | [S12], [S41] | 4 | × | 0 | [S57] | 1 | [S44] | 1 | × | 0 |
| | I | [S84] | 1 | [S91], [S102] | 2 | × | 0 | × | 0 | [S14] | 2 | × | 0 |
| **Modularity** | I | [S39] | 1 | × | 0 | × | 0 | × | 0 | × | 0 | × | 0 |
| **Network Topology** | A | × | 0 | [S15] | 5 | × | 0 | × | 0 | × | 0 | × | 0 |
| **Time Behaviour** | A | [S85] | 1 | [S20], [S99] | 6 | [S85] | 2 | [S19] | 1 | [S20], [S35] | 7 | [S36], [S85] | 3 |
| **Power Consumption** | A | [S90] | 6 | [S99] | 3 | [S90] | 14 | × | 0 | [S80] | 3 | × | 0 |
| **Productivity** | A | × | 0 | [S12], [S38], [S95] | 5 | × | 0 | × | 0 | × | 0 | × | 0 |
| | I | [S5] | 2 | × | 0 | × | 0 | × | 0 | [S14] | 1 | [S5] | 3 |
| **Readability** | A | × | 0 | [S19], [S59], [S98] | 5 | × | 0 | × | 0 | × | 0 | × | 0 |
| **Traceability** | A | × | 0 | [S82] | 2 | × | 0 | × | 0 | × | 0 | × | 0 |
| **Resource Utilization** | A | [S85] | 4 | × | 0 | × | 0 | [S20] | 1 | × | 0 | [S36] | 2 |
| **Reusability** | A | × | 0 | [S3], [S65], [S70], [S96], [S104], [S123], [S139], [S140], [S141], [S142] | 39 | × | 0 | [S3] | 1 | [S3], [S75] | 4 | × | 0 |
| | I | [S45] | 1 | × | 0 | × | 0 | × | 0 | × | 0 | × | 0 |
| **Security** | A | [S89] | 5 | × | 0 | × | 0 | × | 0 | × | 0 | × | 0 |
| | I | × | 0 | [S31] | 1 | × | 0 | × | 0 | × | 0 | × | 0 |
| **Testability** | A | × | 0 | [S3], [S41] | 3 | × | 0 | [S48] | 1 | [S3], [S52], [S75] | 14 | × | 0 |
| | I | [S39] | 2 | × | 0 | × | 0 | × | 0 | × | 0 | | |
| **Understandability** | A | [S23], [S24], [S74] | 13 | [S3], [S59], [S65], [S96], [S98], [S104] ], [S123], [S139], [S140], [S141], [S142] | 38 | × | 0 | [S3], [S121] | 2 | [S3], [S75], [S121] | 7 | [S24] | 1 |
| | I | [S5] | 2 | × | 0 | × | 0 | × | 0 | × | 0 | [S5] | 3 |

*Note for Table 11 and 12: ++PSs, +PSs, --PSs, -PSs represent the PSs with significant positive, positive, significant negative and negative impact on corresponding quality attribute, respectively.*

**Table 13: Comparison between the major findings of this work and Dallal and Abdin [21]**

| RQs | Sub-questions | Dallal and Abdin [21] | | | This SMS | | | | | |
|---|---|---|---|---|---|---|---|---|---|---|
| **Demographic** | How many primary studies were selected? | 76 | | | 142 | | | | | |
| | How the studies were classified according to applied research method? | NI* | | | *Case Study:* 115 | *Experiment:* 18 | *Simulation:* 1 | *Survey:* 1 | *Mixed:* 7 | *Tool:* 11 |
| | Were the selected PSs categorized based on the author setting location? | No | | | Yes | | | | | |
| | How the selected PSs were categorized based on the author's location setting? | *All industry authors:* 2 | *All academic authors:* 65 | *Both:* 9 | *All industry authors:* 5 | *All academic authors:* 122 | | | *Both:* 15 | |
| | Which publication year contributed a maximum number of PSs? | 2011 | | | 2014 | | | | | |
| | How many PSs were published across different publication types? | *Conference proceedings:* 45 | *Journal:* 31 | | *CSW*:* 88 | | | | *JTM*:* 54 | |
| | How many publication venues were identified? | *Total:* 51 | *Conference proceedings:* 33 | *Journal:* 18 | *Total:* 85 | *CSW:* 58 | | | *JTM:* 27 | |
| | Which journal and conference proceeding contributed a large number of PSs? | IST and CSMR | | | IST and ICSME | | | | | |
| **RQ1** | How many refactoring activities were identified? | *Total:* 45 | *Object-oriented:* 9 | *Fowler:* 36 | *Total:* 154 | *Object-oriented:* 84 | | | *Fowler:* 70 | |
| | How many refactoring activities were investigated by more than one PS? | 15 | | | 81 | | | | | |
| | How many studies clearly or partially specified the investigated refactoring activities? | NI | | | *Specific:* 100 | *Partially Specific:* 8 | | | *Unclear:* 34 | |
| | How many studies reported the impact of individual and overall refactoring activities? | *Individual:* 37 | *Overall:* 39 | *Both:* NI | *Individual:* 53 | *Overall:* 74 | | | *Both:* 15 | |
| | What is the classification of selected PSs based on considered refactoring activities? | NI | | | *Fowler catalogue:* 70 | *Object oriented:* 10 | | *Both:* 28 | *Unclear:* 34 | |
| | Among the identified refactoring activities, the impact of how many refactorings is yet to be explored? | 0 out of 45 | | | 52 out of 154 | | | | | |
| **RQ2** | How many software quality measures were identified? | 167 | | | 339 | | | | | |
| | Which quality measures have been used among selected PSs? How many studies used them? | *Internal:* 63 | *Internal as surrogate:* 14 | *External:* 9 | *Internal:* 75 | *Internal as surrogate:* 35 | | | *External:* 52 | |
| | What is the classification of PSs based on software quality measures? | NI | | | *Product:* 89 | *Process:* 24 | | *Resource:* 9 | *Mixed:* 20 | |
| | What is the variety of internal software product quality measures? | *Cohesion:* 36; *Coupling:* 30; *Complexity:* 23; *Size:* 20; *Inheritance:* 7; *Composition:* 2; *Encapsulation:* 4; *Polymorphism:* 1; *Information hiding:* 2 | | | *Cohesion:* 36; *Coupling:* 49; *Complexity:* 44; *Size:* 29; *Inheritance:* 21; *Composition:* 2; *Encapsulation:* 7; *Polymorphism:* 1; *Messaging:* 1 | | | | | |
| | Which metric suites were most frequently employed by the selected PSs? How many studies used them? | NI | | | *MOOD:* 2; *Briand et al.:* 2; *EMOOSE:* 19; *QMOOD:* 14; *L&K:* 7; *MOOSE:* 53; *Others:* 86 | | | | | |
| | What is the maximum number of quality measures used by a PS? | NI | | | 32 | | | | | |
| | Which is the most commonly used software quality measure? | WMC | | | LOC and NOM | | | | | |
| | Is the association of software quality measures to quality attributes in the context of refactoring application is reported? | No | | | Yes | | | | | |

| | Question | | | | | | |
|---|---|---|---|---|---|---|---|
| RQ4 | How many datasets were utilized by the selected PSs? | *Total:* 210 | | *Distinct:* 149 | *Total:* 472 | | *Distinct:* 294 |
| | What is the maximum number of datasets considered by a PS? | 12 | | | 23 | | |
| | How many PSs employed only one dataset? | 35 | | | 66 | | |
| | How many datasets were used by more than two PSs? | 8 | | | 29 | | |
| | Which dataset was considered more frequently? | JHotDraw | | | JHotDraw | | |
| | What is the impact of same refactoring activity applied to a common dataset, on the quality of same quality attribute? | NI | | | Inconsistent | | |
| | How many datasets belong to different sizes, programming languages and dataset types? | *Size:* | *Language:* | *Dataset type:* | *Size:* | *Language:* | **Dataset type:** |
| | | *Total:* 210 | *TD\*:* 149 | *TD:* 149 | *Total:* 472 | *TD:* 294 | *TD:* 294 |
| | | *Small:* 69 | *Java:* 130 | *Open source*: 89 | *Small:* 160 | *Java:* 226 | *Open source:* 192 |
| | | *Medium:* 82 | *C++:* 5 | *Commercial:* 26 | *Medium:* 168 | *C++:* 24 | *Commercial:* 45 |
| | | *Large:* 38 | *C#:* 3 | *Academic:* 28 | *Large:* 87 | *C#:* 20 | *Academic:* 35 |
| | | *NS\*:* 21 | *NS:* 11 | *Student:* 6 | *NS:* 57 | *Python:* 1 | *Student:* 13 |
| | | | | | | *Swift:* 1 | *NS:* 9 |
| | | | | | | *NS:* 22 | |
| | How many studies used datasets pertaining to different sizes, programming languages and dataset types? | *Size:* | *Language:* NI | *Dataset type:* | *Size:* | *Language:* | *Dataset type:* |
| | | *Small:* 32 | | *Only open source*: 39 | *Only small:* 55 | *Only Java:* 115 | *Only open source:* 79 |
| | | *Medium:* 9 | | *Only commercial:* 14 | *Only medium:* 19 | *Only C++:* 10 | *Only commercial:* 21 |
| | | *Large:* 25 | | *Only academic:* 14 | *Only large:* 7 | *Only C#:* 7 | *Only academic:* 21 |
| | | *VS\*:* NI | | *Only student:* 3 | *VS:* 43 | *VL\*:* 3 | *Only student:* 5 |
| | | *NS:* 10 | | *Variable type:* 6 | *NS:* 18 | *NS:* 7 | *Variable type:* 12 |
| | | | | | | | *NS:* 4 |
| RQ5 | How many software quality attributes were identified? | *Total:* 23 | | *Internal:* 10 | *External:* 13 | *Total:* 37 | *Internal:* 10 | *External:* 27 |
| | How many studies measured the impact of refactoring on internal and external quality attributes? | *Internal*: 54 | | *External:* 12 | *Both:* 10 | *Internal*: 62 | *External:* 67 | *Both:* 13 |
| | Do the difference between the outcomes of studies conducted in commercial and industrial settings reported? | No | | | Yes | | |
| | How many PSs applied statistical techniques to assess the effect of refactoring on software quality | 10 | | | 26 | | |
| | How many distinct statistical approaches were considered? | 10 | | | 17 | | |
| | What is the impact of overall refactoring activities on internal software quality among the studies conducted in industrial and academic settings? | NI | | | *Academic setting:* Mostly positive | | *Industrial setting:* Mostly neutral |
| | What is the impact of overall refactoring activities on external software quality among the studies conducted in industrial and academic settings? | NI | | | *Academic setting:* Mostly positive | | *Industrial setting:* Mostly positive |

*\*The full form of the abbreviations used in the Table 11, 12 and 13 are as follows: a) NI: Not Informed (not investigated that particular aspect); b) CSW: Conference/ Symposium/ Workshop ; c) JTM: Journal/ Transaction/ Magazine ; d) TD: Total Distinct; e) NS: Not Specified; f) VS: Variable Sized; g) VL: Variable Language; h) A: Academic Setting; i) I: Industrial Setting*

# 5. DISCUSSION: ISSUES AND RECOMMENDATIONS

The growing interest in the area of determining the impact of refactoring activities on software quality is seeing several vital issues and challenges emerging from the current research. In this section, we discuss various challenges related to finding the effect of refactoring activities on software quality, that are required to be addressed in future.

## 5.1. Refactoring activities (RQ1)

Selected PSs either applied the identified refactoring opportunities to the software, or extracted the applied refactoring activities occurring between two subsequent releases of the software utilizing several approaches mentioned in Section 4.2.1. The techniques may have distinguishable effects on the final outcomes of the underlying experiment as diverse approaches return different information regarding the refactoring activities applied between different releases. To mitigate the impact of such differences, various techniques can be combined together [78]. Thus, researchers are advised to take this finding into consideration while designing an experiment.

The impact of a wide range of refactoring activities on software quality has been analysed in selected PSs. Almost half (45.5%) of the considered refactoring activities belong to Fowler's catalogue [8]. Researchers have shown less interest in working toward determining the effect of other object-oriented (each by five or lesser PSs except *Move Class*) refactoring activities. This may be due to the following reasons:

- The tools used to perform or detect refactoring mostly cover the refactoring activities proposed by Fowler [8]. For example, Ref-Finder [67] is capable of extracting 63 refactoring activities between the two releases of software systems; all belonging to Fowler's catalogue [8].
- It may be due to the popularity and awareness of Fowler's refactoring activities among the developers and researchers.

Therefore, there is a need to accommodate object-oriented refactoring activities other than those defined by Fowler [8] in existing refactoring detection tools. Moreover, an extended repository which also catalogues the details of other object-oriented refactoring activities too, can be established in future.

Among considered refactoring activities, overly half (69.5%) of the refactoring activities were rarely (each by less than five PSs) applied by the selected PSs. Moreover, we noticed that *Domain from Presentation*, and *Convert Procedural Design to Objects* refactoring activities provided by Fowler [8] were not explored by any of the selected PSs. This may be attributed to the unavailability of these refactoring activities in an automated tool, some refactoring activities like *Introduce Indirection* are not frequently used in industry [79], or some refactoring activities like *Substitute Algorithm* cannot be automated at all [80], resulting in a difficulty to analyse the presence and impact of such refactorings. The analysis of the impact of less or unexplored refactoring activities on the quality of software systems is an open research area.

About two third (62.7%) of the PSs reported the effect of overall refactoring activities on software quality. Comparatively, a less (42.3%) proportion of studies classified each refactoring activity according to their measurable impact on software quality. But, these PSs did not report the impact of 52 (out of 154) individual refactoring activities on any of the software quality attributes. The key reason behind this may be due to the inability of existing refactoring detection tools like Ref-Finder [67] or, refactoring identification and application tools like JDeodorant [81] to extract or, identify and apply other (not proposed by Fowler [8]) object-oriented refactoring activities, respectively. Even for remaining 102 refactoring activities, these PSs do not cover all software quality attributes. Hence, there is a need to investigate the effect of each refactoring activity on individual quality attributes. This classification will assist the software developers in achieving their design objectives by selecting most beneficial refactoring activities with respect to specific software quality attributes. Also, about one third (29.6%) of the PSs have not clearly stated the considered refactoring activities. Researchers are suggested to properly mention the selected refactoring activities while reporting their works. Otherwise, it will be difficult to determine which refactoring activities are affecting the quality attributes in a negative or positive way.

Most (134 PSs) of the selected PSs have not provided any valid justification whilst choosing the refactoring activities for their studies. Based on previous studies [79, 82], refactoring activities that the developers most or least frequently perform as part of their daily maintenance tasks, were selected by a very limited number of ([S30], [S43], [S50], [S82], [S93], [S141] and [S142]) studies. Kannangara and Wijayanayake [S36] have chosen the refactoring activities from the rankings provided by previous studies [S58]. Consequently, the most commonly used refactoring activities in industry [79, 83] like *Rename Field*, *Rename Class*, *Rename Package*, *Extract Local Variable*, etc. received less attention by PSs. This observation indicates the gap between academic and industry research in terms of refactoring activities exercised by academic researchers and refactoring activities actually applied in practice by industry practitioners. Hence, researchers are suggested to involve industry professionals or extract the related information from previous studies [79, 82-83] to select most frequently used refactoring activities by developers rather than considering rarely applied refactoring activities like *Extract Class* (39 PSs), *Extract Superclass* (18 PSs) and *Introduce Parameter Object* (10 PSs). The analysis of the impact of these more applied refactoring activities on the quality of software will produce more productive results for the software industry. Furthermore, it would be advantageous if the experiments using these refactoring activities are performed on commercial software systems in industrial settings.

---

***Issues and recommendations***

***Issue 1:*** *The authors of only few PSs involved industry practitioners or considered the rankings provided by previous studies whilst selecting the refactoring activities. As a result, many of the most commonly used refactoring activities like Rename Field and Extract Local Variable have received less attention by the selected PSs.*

***Recommendation1:*** *It is advised to consider such aforementioned factors while choosing refactoring activities as analysis of the impact of these refactoring activities on software quality will produce more productive results for the industry practitioners.*

***Issue 2:*** *The authors of many studies failed to mention the specific refactoring activities considered in their research. Also, a limited number of studies provided the classification of refactoring activities based on their impact on desired quality attributes.*

***Recommendation 2:*** *Researchers are suggested to further conduct the empirical studies by investigating the impact of each specific refactoring activity on individual quality attributes rather than on software quality in general.*

---

*5.2. Software quality measures (RQ2)*

Most of the works used internal quality measures as a means to assess the impact of refactoring activities on external quality attributes. Investigating the refactoring effect on external quality attributes by directly using external quality measures has received less attention (36.6% PSs). This may be due to the high effort required for directly measuring such quality attributes, as these measures cannot be obtained directly from the source code or when the software is deployed. Rather, one has to do a thorough analysis of maintenance data stored in software repositories over a considerable duration of time in order to acquire the value of external measures. The future studies should utilize external measures while investigating the impact of refactoring activities on external quality attributes in order to provide more reliable results, and to validate the findings resulting from using internal quality measures as surrogates. Moreover, among the internal quality measures, it is observed that cohesion, coupling, size, complexity and inheritance measures received more attention across the selected PSs, in comparison to other quality measures like polymorphism, data encapsulation, information hiding, etc. The frequent usage of these measures may be attributed to the fact that researchers believe these internal measures are more strongly related to software quality. As a result, they consider them as benchmarks for measuring internal quality or potentially relevant in terms of combining them together for measuring external quality. The another reason may include the more availability of metrics that quantify these measures. However, the less explored measures are also the basic pillars of object-oriented paradigms and are an important aspect to explore. Hence, to enrich the body of knowledge, there is a need to investigate the impact of refactoring activities on software quality using the unexplored set of other measures too.

A large variety of quality measures have been employed by the selected PSs. For instance, a number of cohesion measures like TCC, LCOM, LSCC, etc. were used to measure the cohesion among code artifacts. These cohesion measures may conflict with each other on the application of refactoring activities, i.e. the improvement of one cohesion measure may cause another measure to degrade [55]. Therefore, a study using one cohesion measure may find an improvement in a quality attribute, whereas another work utilizing a different cohesion measure may experience the abatement in the same quality attribute, following the application of the same refactoring activity. So, the researchers are encouraged to propose a standard unified metric framework, which will provide more significant measure of the software quality on a *relative* metric scale.

> **Issues and recommendations**
> **Issue 1:** A large number of PSs determined the effect of refactoring activities on external quality attributes by using internal quality measures.
> **Recommendation 1:** Researchers are suggested to use direct measures of external software quality to support the conclusion provided by utilizing internal measures as proxies.
> **Issue 2:** Most of the studies utilized cohesion, coupling, size, complexity and inheritance metrics to measure the impact of refactoring on software quality. Many other internal quality measures like polymorphism, data encapsulation, information hiding etc. were considered infrequently by the researchers.
> **Recommendation 2:** It is advised to investigate the impact of refactoring activities on software quality attributes utilizing less explored software quality measures.
> **Issue 3:** A PS utilizing one cohesion measure may find an improvement in a quality attribute, whereas another work utilizing a different cohesion measure may experience the abatement in the same quality attribute, following the application of the same refactoring activity.
> **Recommendation 3:** It is suggested to propose a standard unified metric framework, which will provide more significant measure of the software quality on a relative metric scale.

### 5.3. Refactoring impact prediction tool (RQ3)

Among 142 PSs, only eleven PSs contributed software tools to measure the effect of refactoring activities on software quality. But, these tools address only few distinct (23 out of 154) refactoring activities, and (15 out of 37) software quality attributes. Moreover, most of these tools ([S34], [S56], [S88], [S94] and [S104]) are either obsolete or not publicly available. This unavailability of tools, which can predict the effect of changes caused by each refactoring activity or estimate the effort required to perform refactoring activities, discourage developers from adopting refactoring in practice. Furthermore, software developers want to assess only the changed parts of the source code. Although, there exist many refactoring tools that identify refactoring opportunities or perform refactoring activities, these tools cannot run only on the differential source code [16]. This led us to the conclusion that development of the updated tools to predict or assess the actual benefits of refactoring activities is an open research area.

The tools provided by [S70] and [S104] address only five distinct quality attributes namely reusability flexibility, extensibility, effectiveness and understandibility. The other tools estimate the impact of refactoring activities on only internal quality attributes. Many managers and developers are reluctant to apply refactoring activities because of the possibility of introducing bugs, and cost/effort required to make a change in the code [16]. Instead the mapping of the impact of refactoring activities on internal quality to directly measurable external quality attributes like defect proneness, productivity, cost, effort, etc., will be more beneficial for the industry developers. Accordingly, researchers are encouraged to extend the existing tools by including these directly relevant external software quality attributes. This will further assist the developers in determining whether to apply a certain refactoring activity for the improvement of a desirable quality attribute.

In [S34], the authors validated the proposed tool utilizing a small student project written in Java. In [S56], researchers evaluated the tool on a C++ project in academic setting, but did not report the project size. *AutoRefactoring* [S70], *OBEY* [S88], *MultiRefactor* [S92], *CODe-Imp* [S94, S104], *JDeodorant* [S118, S130, S131] and *Refactoring Scheduler* [S121] tools were evaluated on Java datasets in academic settings. Researchers are advised to validate the tools with experienced industry developers on large industrial as well as open source projects representing a variety of domains. Further, these existing tools estimate or predict the effect of refactoring activities on the quality of software systems written in Java and C++ programming languages only. Hence, the development of new tools addressing the impact of refactoring on the quality of other object-oriented software written using languages like Smalltalk, C#, Python, etc. is recommended to meet the current industrial needs and trends.

An Integrated Development Environment (IDE) could play a vital role in the adoption of refactoring tool support [83]. These days, almost all the most popular IDEs available provide a considerable support for refactoring. But, these IDEs do not generally assist in measuring or predicting the effects of refactoring activities on software quality, which results in one of the major roadblocks in adopting refactoring practices in software industry [16, 84]. Therefore, researchers are suggested to develop IDE plugins for estimating the impact of applying specific refactoring activities on quality of the software under development. Appending refactoring benefits information with automatic refactoring support can assist the schedule-bound development teams in better understanding the refactoring benefits, and also in performing product negotiations with customers during software acceptance.

Also, the prediction of refactoring impact on software quality before applying the refactoring activities can potentially save a lot of time and effort. This valuable time and effort would otherwise be wasted in implementing inappropriate refactoring activities as well as in rolling back from such refactoring activities that degrade software quality [85]. Nonetheless, Kim et al. [84] summarizes this need as: *"I'd love a tool that could estimate the benefits of refactoring"*.

> ***Issues and recommendations***
> ***Issue 1:*** *There are only eleven software tools that estimate the effect of software refactoring activities on software quality. Majority of these tools are either obsolete or unavailable. Additionally, these tools do not address all major refactoring activities and software quality attributes.*
> ***Recommendation 1:*** *Development of the updated tools employing more refactoring activities and software quality attributes, to quantitatively predict/assess the actual effect of each refactoring activity on various software quality attributes is an open research area.*
> ***Issue 2:*** *Most of the existing tools predict or assess the impact of refactoring activities only on Java source code.*
> ***Recommendation 2:*** *It is advised to develop such tools that address the impact of refactoring on the quality of datasets written using less explored object-oriented languages like Smalltalk, C#, Python, etc. to meet the current industrial trends.*

### 5.4. Datasets used (RQ4)

The software systems used for conducting empirical studies play a key role in establishing performance benchmarks. These benchmarks can be later used by other researchers working in the related areas as reliable sources to compare their research with existing research. Many datasets have been utilized frequently to empirically evaluate the selected set of PSs. This distribution will assist the researchers in identifying the most commonly used datasets for performing their empirical studies. JHotDraw, GanttProject and Apache Ant have been extensively used as datasets in PSs. Also, Apache Ant and its different versions have been used in recent studies to study the impact of refactoring activities on software quality. This may be due to the following reasons.

- Many of the studies employing these extensively used datasets are conducted by the same researchers, hence some of the related data (e.g. information regarding refactoring activities applied on the dataset, software quality measures, etc.) was already available to them for performing their recent work.
- These datasets are large in size and are well known for their good design.

The major population of selected PSs (51.4%) have either used only small datasets, or failed to report the size of the dataset. The evaluation of empirical studies using large datasets can generate more trustworthy conclusions [21, 25]. Moreover, the recent (2011 onwards) papers started involving medium as well as large datasets along with smaller datasets in almost equal proportion. Hence, as the research domain is maturing, researchers are generating more reliable conclusions employing the datasets with variety of sizes.

The related research has predominantly been performed on the datasets written in Java language, followed by C++ and C# languages. This may be due to either Java's popularity (the examples provided in Fowler's book are in Java), or the lack of availability of tools for other programming languages. There is a need to validate the impact of refactoring activities on software quality by utilizing the datasets written in other object-oriented languages like Smalltalk, Python, etc. Furthermore, we did not report the variations among the outcomes of selected PSs resulting after employing different versions of Java datasets as most (92.3%) of the PSs did not provide any information regarding the same. The difference in versions of datasets may play a key role as well towards establishing a better insight regarding the benefits of refactoring with the change of dataset versions. Therefore, researchers are advised to mention the versions of employed Java datasets. In addition, empirical studies can be performed to investigate the change in impact of refactoring on software quality following the incorporation of datasets with different versions. About 66 (46.5%) PSs used a single dataset for empirically evaluating the effect of refactoring activities on software quality. Out of these 66 PSs, the size of datasets used in 78.8% of the PSs was either small or not specified. Moreover, only three ([S11, S78, S90]) PSs utilized datasets written in different programming languages. Hence in order to boost the generalization of results and study the trends with the change of dataset characteristics, this area can be further explored to perform empirical studies by utilizing the software systems pertaining to different programming languages, platforms, sizes and domains.

Numerous open source (76.1%) datasets have been commonly used to evaluate the impact of refactoring activities on software quality. These open source datasets were employed only by the PSs performed in academic settings, whereas studies conducted in industrial settings utilized only commercial datasets. Open source projects are clearly preferred in performing comparative studies and for enabling repeatability of studies. But the differences among the outcomes (regarding the effect of refactoring on software quality) of studies performed in academic and industrial settings might be due to employed dataset types. Therefore, researchers are also encouraged to use commercial systems more often to further validate their results and arrive at a generalizable conclusion. Some researchers also used academic or student datasets instead of publicly available or commercial datasets. The use of these datasets becomes less useful from the viewpoint of research replication.

Few PS groups (e.g. [S7, S54, S68], [S38, S68], etc.) showed contradictory evidences on the impact of same refactoring activities applied to the same datasets. For example, Ratzinger et al. [S54] found that the number of defects in the preceding time period decreases with the increase in the number of refactoring activities whereas an opposite trend is witnessed by Weißgerber and Diehl [S68], for ArgoUML dataset. Therefore, the area demands future work towards deriving trustworthy conclusions after resolving these contradicting evidences.

---

***Issues and recommendations***

***Issue 1:*** *Most of the researchers performed case studies in an academic settings using undergraduate students with no or little industry experience. These studies are hence not reliable in understanding whether the industry developers take software quality as an important factor while applying a refactoring activity or care about the effect of refactoring activities on software quality.*

***Recommendation 1:*** *It is advised to empirically validate the studies in commercial setting involving experienced developers, industrial datasets, and commonly applied refactoring activities for more realistic findings.*

***Issue 2:*** *Very few PSs utilized the datasets belonging to more than one programming language, type or size while empirically investigating the effect of refactoring on software quality.*

***Recommendation 2:*** *Researchers are suggested to conduct further research by employing datasets representing different domains, programming languages, types, and sizes to study the trends of change in dataset characteristics.*

***Issue 3:*** *PSs investigating the effect of same refactoring activities employing same datasets reported contradictory evidences.*

***Recommendation 3:*** *Hence, the area can be further explored to generate reliable conclusions after resolving these inconsistent findings.*

---

### 5.5. *Refactoring impact on quality (RQ5)*

A large number of internal (ten) and external (27) quality attributes are taken into account among selected PSs. We observed that cohesion, coupling, size, complexity and inheritance attributes were investigated more frequently than any other quality attributes. Furthermore, each of almost 50% of internal and 78% of external quality attributes found in selected PSs were addressed by a very limited set of PSs (eight or less). The reason behind preferring internal quality attributes may include simplicity in calculating the effect of refactoring on internal quality attributes, or these quality attributes are considered as key quality attributes from the user or practitioners' point of view. But, several quality attributes like security, readability, extensibility [86], etc., among this substantial proportion, are considered more important by industry practitioners. Therefore, more empirical studies are required by considering such less explored software quality attributes in order to strengthen the relevant evidence regarding refactoring effect associated with such attributes. Furthermore, among the studies focusing on determining the impact of refactoring activities on external quality attributes, each of nearly 50% of studies concentrated on only single quality attribute. Though a considerably large task, in order to understand the actual impact of refactoring on software quality 'as a whole', it is suggested to consider a comprehensive set of ISO quality attributes [77] or other widely accepted quality models (McCall's [87], Boehm's [88], etc.) that are well-suited for object-oriented software. Another feasible direction could be a kind of meta-analysis based on quantitative study, which statistically combines the outcomes of the studies targeting individual quality attributes.

The impact of about 89% of refactoring activities on four internal and 20 external (including industry relevant [86]) quality attributes has been addressed infrequently. This massive (89%) proportion also includes many refactoring activities like *Rename Field*, *Rename Class*, *Rename Package*, *Extract Local Variable,* etc., which are considered as most important in industry [79, 83]. Rather, researchers have mainly emphasized on determining the impact of *Move Method, Extract Method* and *Extract Class* refactoring activities on cohesion, coupling, complexity and size attributes. This may signify that researchers regard these refactoring activities as more promising for software quality improvement in comparison to other refactoring activities. As a result, they are showing more interest towards proving or validating the aforementioned assumptions by conducting more empirical studies. Therefore, researchers are advised to also investigate the influence of less considered but industry relevant refactoring activities on least explored (but used in industry) quality attributes while designing experiments.

Only a small proportion (26 PSs) of selected PSs determined the statistical significance of results regarding the impact of refactoring activities on software quality attributes. Due to such less number of PSs, we were not able to combine the reported outcomes based on p-values. Hence, we were only left with vote-counting approach, which may be erroneous [89]. Therefore, it is imperative to explore the statistical significance of results concerning the effect of refactoring on software quality in order to arrive at more reliable conclusions. Furthermore, among these 26 PSs, 10 PSs were found to be overlapping with the statistically significant PSs selected by Dallal and Abdin [21]. The additional 16 PSs assisted us in drawing more reliable conclusions.

Except for the quality attributes having no significant votes, we considered only significant votes while calculating the final impact of refactoring on software quality, as also mentioned in Section 4.2.5. The number of non-significant votes were much higher in number than significant vote results because a majority (82%) of PSs did not apply any statistical techniques. As the non-significant votes are unreliable, a huge proportion of these non-significant votes were ignored for quality attributes holding significant votes. It is advised to report more reliable conclusions based on robust statistical techniques, while evaluating the effect of refactoring on software quality. While applying the aforementioned approach, the calculated final impact of refactoring on software quality may change for some of the software quality attributes, upon consideration of non-significant votes along with significant vote-counting results. For example, the number of *positive*, *significant positive*, *negative*, *significant negative, unchanged* and *insignificantly changed* votes for the impact of *Extract Class* refactoring on cohesion are 36, zero, four, five, zero and zero, respectively. If all vote-counting results are taken into consideration, the final impact of *Extract Class* refactoring on inheritance will be positive, which is calculated as negative (in Figure 22) after ignoring the non-significant votes. This area can be further explored by deriving reliable results after checking the statistical significance of already conducted non-significant empirical studies as these results will also play an important role in establishing the existing knowledge regarding the effect of refactoring on software quality.

The results obtained from the studies conducted in academic settings do not match with the findings acquired from the studies performed in industrial settings. For academic settings, the application of overall refactoring activities resulted in improvement of internal quality attributes. On the contrary, industry practitioners did not find any definitive impact of refactoring activities on internal software quality attributes. Similarly, dissimilar trends are observed for the studies investigating the effect of refactoring activities on external software quality attributes. The difference regarding the impact of refactoring activities (individual and overall) on software quality among the findings of studies performed in academic and industrial settings may be due to following reasons:

a) The number of studies conducted in industrial settings (17 PSs) are much less than the studies performed in academic settings (125 PSs).
b) The type of datasets used among the studies conducted in these settings are different. The industrial studies utilized only commercial datasets, whereas a majority of the academic studies employed only open source systems.
c) Approximately half of the industrial studies used different software quality measures as compared to those used in academic studies. Software quality measures also play an important role, as a study employing one quality measure may result in the abatement of a quality attribute, whereas the another study using different quality measures may experience an improvement in the same quality attribute [54].

Therefore, there is a need to validate the studies in both academic and industrial settings in order to resolve the inconsistent findings. Further to that, the findings of this SMS are identical to Dallal and Abdin [21] only for one internal and eight external quality attributes. This may be attributed to the difference in either total number of PSs or employed inclusion/exclusion criteria. Also, the influence of refactoring activities on three internal and nearly 14 external quality attributes has not been validated in industrial setting. But many of these quality attributes like readability, adaptability [86], etc. are of more interest for industry practitioners and hence require further investigation in industrial settings.

There are total 89 PSs which reported the effect of overall refactoring activities on software quality attributes. These PSs present different results for the effect of overall refactoring on internal/external quality attributes. Further, it is observed that the application of overall refactoring activities affected several software quality attributes in an inconsistent manner. Thus, there is further need to investigate the impact of refactoring in general on such software quality attributes. Additionally, from such PSs, it is difficult to determine about which of the refactoring activities affect the software quality attributes in positive ways. For example, if the application of *Extract Class* and *Inline Method* refactoring activities resulted in negative software maintainability, both the refactoring activities cannot be held equally responsible for this effect. Therefore, some (60) PSs also reported the effect of 102 individual refactoring activities on internal/external quality attributes in order to clarify this notion. But this classification does not cover the impact of each refactoring activity on every quality attribute. Moreover, the conclusions of these PSs are also inconsistent in terms of the effect of a specific refactoring activity on same quality attribute or multiple quality attributes. For example, in a PS [S65], *Extract Method* improves cohesion whereas in another PSs [S25], it deteriorates software cohesion. Therefore, the area of determining the effect of individual as well as overall refactoring activities on software quality attributes demands further investigation to draw definite conclusions.

The reasons behind inconsistent or diverging results concerning the impact of overall or individual refactoring on software quality may include the following:

a) While assessing the impact of overall refactoring activities on software quality, the number and type of refactoring activities as well as their sequences of application on source code are different among selected PSs.
b) Researchers opted different approaches to gather information regarding the refactoring activities from software repositories. These techniques return different results about the refactoring activities applied between different releases [78].
c) The usage of different datasets in terms of dataset type, language and size among selected PSs.
d) The selected PSs employed several different quality measures which may quantify different aspects of a particular quality attribute following different definitions for the same.
e) The variation in considered refactoring opportunity identification techniques as some of these approaches are error prone and may identify wrong refactoring opportunities. As a result, the application of a refactoring activity on software code which did not need refactoring in actual, may cause the quality to degrade.
f) The exploration of dissimilar refactoring application approaches or tools among selected PSs because these techniques may produce disparate refactored piece of code. Also, the existing tools are error-prone [84] and may generate incorrect code, which may also negatively impact software quality.

> ***Issues and Recommendations***
>
> ***Issue 1:*** *The findings obtained from the studies conducted in industrial settings are not consistent with the ones conducted in academic settings.*
> ***Recommendation 1:*** *It is advised to validate the studies in both academic and industrial settings in order to resolve the inconsistent findings.*
> ***Issue 2:*** *Several refactoring activities proposed by Fowler either remained unexplored or are considered infrequently. Furthermore, the selected PSs considered few internal and external quality attributes. Consequently, many important quality attributes like readability, security, etc. have received less researchers' attention.*
> ***Recommendation 2:*** *Researchers are encouraged to consider scarcely explored refactoring activities, and quality attributes by keeping in mind ISO quality attributes or other accepted quality models.*
> ***Issue 3:*** *A large number PSs ignored the exploration of statistical significance of results concerning the effect of refactoring on software quality.*
> ***Recommendation 3:*** *It is suggested to report more statistically significant conclusions by supporting their empirical studies with statistical techniques.*
> ***Issue 4:*** *The results of selected PSs regarding the effect of software refactoring activities on software quality are contradictory because different refactoring activities affect different software quality attributes in a variety of ways.*
> ***Recommendation 4:*** *Investigation of the impact of individual as well as overall refactoring activities on software quality attributes is still an open research issue and needs further in-depth exploration.*

### *5.6. Comparison of Findings with Dallal and Abdin [21]*

Although this SMS is different from Dallal and Abdin's SLR [21] in several aspects as discussed in Section 2.2. However, because of the similarity of the research questions, we still decided to compare the outcomes of our work with that of Dallal and Abdin [21]. Table 13 summarizes the comparison of the major findings of common research questions.

The mentioned differences between the results of our work and aforementioned SLR [21] can be attributed to the difference in number as well as nature of selected PSs and our defined inclusion/exclusion criteria. Given the differences, it is on the wisdom of the reader to infer the appropriate findings depending upon their scope of interest.

## 6. THREATS TO VALIDITY

The main threats to the validity of our SMS arise from the unintended biases during the search of relevant literature, primary study selection, quality assessment procedure, and data extraction from the selected studies. We followed the classification of threats provided by Wohlin et al. [90] which includes conclusion, construct, internal, and external threats to discuss the aforementioned limitations along with the countermeasures taken to minimize them.

### 6.1. Conclusion validity

Conclusion validity refers to reproducibility, which allows other researchers to yield same results as that of original study on replication [91]. To ensure repetition, the search string, electronic databases for automatic search, publication venues for manual search, inclusion and exclusion criteria, and detailed systematic approach followed to carry out this study, have been clearly stated in this study. In addition, all the data related to the whole SMS process is available online [36]. However, dissimilar results might be produced by other researchers due to their different perceptions when dealing with the study selection phase. To mitigate this threat, the titles, abstracts and full texts of the selected articles were screened independently by both the authors considering the defined inclusion and exclusion criteria. Furthermore, in case of disagreements, discussions were conducted between the authors to resolve discrepancies observed during study selection process.

### 6.2. Construct validity

Construct validity is related to the search and selection of primary studies. We followed a well-defined protocol defined by Kitchenham and Charters [31] to identify the primary studies in an unbiased manner. Firstly, a specific search string was constructed to select studies specific to our phenomenon of interest. We further performed the keyword validation of our search terms against 20 reference articles to refine our initial set of keywords, or to identify additional search terms. The automatic search for the research articles was conducted in most appropriate and renowned electronic databases as suggested by Dyba et al. [39], and Kitchenham and Charters [31]. However, there may be the chance of missing some relevant literature due to the keywords used to search the publication databases. To mitigate the challenges of the automatic search phase, and to validate our search string, a pilot study was performed by both the authors on a total of 20 reference articles. The results of the pilot study favoured our search string formulation. The automatic search process was later complemented with manual search on top ranked journals and conference proceedings specific to software refactoring domain to identify any relevant studies that we might miss out. The references of the selected studies and previous surveys were reviewed independently by both the authors to improve the reliability of our search process and to further minimize the validity threat of missing relevant primary studies. The study selection phase includes inclusion/exclusion criteria, and quality assessment process. Despite this thoroughness, we might have missed some significant articles due to article inaccessibility or language barrier.

### 6.3. Internal validity

The screening of studies using inclusion and exclusion criteria along with the quality assessment process are vulnerable to reviewer's bias in excluding any significant studies. However, utmost care was taken to defend against these threats. To clarify any ambiguous aspects of inclusion and exclusion criteria, a random selection of 90 papers obtained during the automatic search were screened by both the authors independently. The results obtained after screening were matched and discrepancies were resolved. The first author performed the quality assessment and data extraction of the primary studies, and assigned the results to the second author for cross-checking each decision, as suggested by Staples and Niazi [64]. Furthermore, we adopted the well-defined quality assessment criteria recommended by Kitchenham and Charters [31] to exclude the studies that do not meet standards defined for this study. Both the authors were fully involved in study classification process, and any differences in the opinion were resolved through discussion to ensure mutual conformity.

### 6.4. External validity

This study primarily targets the advantages and disadvantages of refactoring activities in object-oriented software arena. The results obtained related to the impact of refactoring activities on the quality may vary across different programming paradigms as there exist a number of dissimilarities among object-oriented software and non object-oriented software systems. Also, it is noteworthy that the findings of this SMS are mainly applicable for the refactoring activities used to remove code smells. The results of this study may differ for the refactoring activities applied to other software artifacts such as requirement specifications, design, etc. This study may serve as a starting point for many researchers to further identify and classify the impact of refactoring on the quality of object-oriented software.

# 7. CONCLUSIONS AND FUTURE WORK

This paper reports a profound systematic mapping study that identifies, assesses, and presents the available empirical literature concerning the impact of refactoring activities on software quality, with the aim of specifying the current state of the art along with the identification of potential open challenges. Firstly, we provided a brief background on software refactoring process. Later, we discussed the mapping study protocol followed to conduct this SMS.

A two-phase approach is followed to search the relevant literature related to our phenomenon of interest. We initially performed an automatic search in six electronic databases namely IEEE Xplore, ACM Digital Library, Springer, ScienceDirect, Scopus, and Wiley. Thereafter manual search was conducted in each of the top 5 renowned journals and conference proceedings. In addition, we performed reference snowballing to increase the reliability of our search and mitigate the risk of missing any relevant literature. Out of 13,283 search results, 142 studies were placed into the final pool of PSs after careful screening against our study selection and quality assessment criteria. These PSs were further classified and examined to answer our formulated research questions.

The mapping of PSs with respect to publication years indicated that determining the effect of refactoring activities on software quality is a highly active area of research. The results of this SMS show that *Move Method, Extract Method*, and *Extract Class* refactoring activities have been frequently investigated, and are of greater interest to researchers than other refactoring activities. We observed that MOOSE metric suite has received considerably more researchers' attention than other software quality measures in order to either determine the effect of refactoring on internal quality attributes, or as surrogates to measure the refactoring impact on external quality attributes. Researchers have paid less attention toward developing automated tool to predict the benefits of software refactoring before its actual application. Numerous different sized datasets implemented in several object-oriented programming languages were used to empirically evaluate the effect of refactoring activities on software quality. Most of the PSs used open source software systems as datasets, which assists the repeatability of studies.

We applied vote-counting approach while combining the findings concerning the effect of refactoring on software quality because only a few PSs explored the significance of reported outcomes by employing the statistical techniques. It is observed that software refactoring does not always improve all software quality attributes as the impact of overall refactoring activities on different software quality attributes yields contradicting evidences. Similarly, different refactoring activities also have opposite effects on different software quality attributes. Finally, we observed that a variety of dimensions like sequence of application of refactoring activities, software quality measure, etc., are involved in studying the effect of refactoring on software quality. The variation of each dimension can impact the final outcome of the study. Hence, each dimension should be handled carefully while designing an experiment.

We expect that this work will assist the industry practitioners in procuring a detailed understanding of the current state of research related to determining the impact of refactoring activities on software quality. Moreover, the limitations and challenges discussed in this SMS will facilitate the researchers in identifying those critical research gaps which require more attention in future.

# APPENDIX

**PRIMARY STUDIES**